\documentclass[a4paper,12pt]{article}
\usepackage{amsfonts,amsthm,amsmath,amssymb,graphicx,color}
\usepackage{hyperref}
\usepackage[lmargin=3.0cm, rmargin=1.5cm,tmargin=2.50cm,bmargin=2.50cm]{
geometry}
\usepackage{mathrsfs}
\usepackage{float}
\usepackage{ulem}
\usepackage{graphicx}

\newcommand{\be}{\begin{equation}}
\newcommand{\ee}{\end{equation}}
\newcommand{\bea}{\begin{eqnarray}}
\newcommand{\eea}{\end{eqnarray}}

\newcommand{\nn}{{\nonumber\\}}
\newcommand{\Tr}{\text{Tr}}

\newcommand\fnsep{\textsuperscript{,}}

\newcommand\beq{\begin{equation}}
\newcommand\eeq{\end{equation}}

\renewcommand{\thefootnote}{\fnsymbol{footnote}}

\begin{document}


\def\gap#1{\vspace{#1 ex}}
\def\be{\begin{equation}}
\def\ee{\end{equation}}
\def\bal{\begin{array}{l}}
\def\ba#1{\begin{array}{#1}}  
\def\ea{\end{array}}
\def\bea{\begin{eqnarray}}
\def\eea{\end{eqnarray}}
\def\beas{\begin{eqnarray*}}
\def\eeas{\end{eqnarray*}}
\def\del{\partial}
\def\eq#1{(\ref{#1})}
\def\fig#1{Fig \ref{#1}} 
\def\re#1{{\bf #1}}
\def\bull{$\bullet$}
\def\nn{\\\nonumber}
\def\ub{\underbar}
\def\nl{\hfill\break}
\def\ni{\noindent}
\def\bibi{\bibitem}
\def\ket{\rangle}
\def\bra{\langle}
\def\vev#1{\langle #1 \rangle} 
\def\lsim{\stackrel{<}{\sim}}
\def\gsim{\stackrel{>}{\sim}}
\def\mattwo#1#2#3#4{\left(
\begin{array}{cc}#1&#2\\#3&#4\end{array}\right)} 
\def\tgen#1{T^{#1}}
\def\half{\frac12}
\def\floor#1{{\lfloor #1 \rfloor}}
\def\ceil#1{{\lceil #1 \rceil}}

\def\mysec#1{\gap1\ni{\bf #1}\gap1}
\def\mycap#1{\begin{quote}{\footnotesize #1}\end{quote}}
\def\ubsec#1{\gap1\ni\underbar{#1}\gap1}

\def\Om{\Omega}
\def\a{\alpha}
\def\b{\beta}
\def\l{\lambda}
\def\m{\mu}
\def\n{\nu}
\def\om{\omega}
\def\s{\sigma}
\def\g{\gamma}
\def\t{\tau}
\def\k{\kappa}

\def\lan{\langle}
\def\ran{\rangle}
\def\f{\frac}

\def\bit{\begin{item}}
\def\eit{\end{item}}
\def\benu{\begin{enumerate}}
\def\eenu{\end{enumerate}}
\def\eps{\epsilon}


\def\bT{{\bar T}}
\def\bL{{\bar L}}
\def\lt{{\tilde \lambda}}
\def\vt{{\tilde v}}
\def\wt{{\tilde w}}
\def\omt{{\tilde\omega}}
\def\ut{{\tilde u}}
\def\bP{{\bf P}}
\def\bQ{{\bf Q}}
\def\Jt{{\tilde J}}
\def\zb{{\bar z}}
\def\wb{{\bar w}}
\def\hb{{\bar h}}
\def\nor#1{{:\kern-1pt #1 \kern-1pt:}}
\def\wc{{\mathcal W}}
\def\sech{{\rm sech}}
\def\csch{{\rm cosech}}
\def\tx{{\tt x}}



\rightline{TIFR/TH/15-36}
\vspace{1.2truecm}

\vspace{1pt}


\begin{center}{\large\bf Thermalization in 2D critical quench and UV/IR mixing}
\end{center}

\vskip.9cm

\thispagestyle{empty} \centerline{\bf Gautam
  Mandal\footnote{mandal@theory.tifr.res.in}, Shruti
  Paranjape\footnote{shrutip@students.iiserpune.ac.in}\footnote{Visiting
    student from IISER, Pune}, and Nilakash
  Sorokhaibam\footnote{nilakashs@theory.tifr.res.in}}
    
\vspace{.8cm} 
\centerline{{\it Department of Theoretical Physics}}
\centerline{{\it Tata Institute of Fundamental Research, Mumbai
    400005, India.} }


\gap7

\centerline{\today}

\gap3

\thispagestyle{empty}

\gap6

\renewcommand*{\thefootnote}{\arabic{footnote}}
\setcounter{footnote}{0}

\centerline{\bf Abstract}
\vskip.5cm 

We consider quantum quenches in models of free scalars and fermions
with a generic time-dependent mass $m(t)$ that goes from $m_0$ to
zero. We prove that, as anticipated in MSS \cite{Mandal:2015jla}, the
post-quench dynamics can be described in terms of a state of the
generalized Calabrese-Cardy form $|\psi \rangle$= $\exp[-\k_2 H
  -\sum_{n>2}^\infty \kappa_n W_n]| \hbox{Bd} \rangle$.  The $W_n$
($n=2,3,...$, $W_2=H$) here represent the conserved $W_\infty$ charges
and $| \hbox{Bd} \rangle$ represents a conformal boundary state.  Our
result holds irrespective of whether the pre-quench state is a ground
state or a squeezed state, and is proved without recourse to
perturbation expansion in the $\kappa_n$'s as in MSS. We compute exact
time-dependent correlators for some specific quench protocols $m(t)$.
The correlators explicitly show thermalization to a generalized Gibbs
ensemble (GGE), with inverse temperature $\b= 4\k_2$, and chemical
potentials $\mu_n=4\k_n$.  In case the pre-quench state is a ground
state, it is possible to retrieve the exact quench protocol $m(t)$
from the final GGE, by an application of inverse scattering
techniques. Another notable result, which we interpret as a UV/IR
mixing, is that the long distance and long time (IR) behaviour of some
correlators depends crucially on {\it all} $\kappa_n$'s, although they
are highly {\it irrelevant} couplings in the usual RG parlance. This
indicates subtleties in RG arguments when applied to non-equilibrium
dynamics.

\setcounter{page}{0} \setcounter{tocdepth}{2}

\newpage

\tableofcontents

\section{\label{sec:intro}Introduction and Summary}

The dynamics of systems undergoing a quantum quench has been
extensively studied in recent years \cite{Polkovnikov:2010yn}. In a
quantum quench, some parameter of the Hamiltonian changes over a brief
period of time. The initial wavefunction in the pre-quench phase,
whether it is a ground state or otherwise, typically evolves to a
non-stationary state, which then evolves by the post-quench
Hamiltonian which is time-independent. An important question in such
dynamics is whether correlators equilibrate at long times, and if so,
whether the equilibrium is described by a thermal ensemble or
otherwise \cite{Polkovnikov:2010yn,Nandkishore:2014kca,
  gogolin2015equilibration}.  With the advent of AdS/CFT, the issue of
thermalization has assumed additional significance as it maps to the
subject of gravitational collapse to a black hole
\cite{Bhattacharyya:2009uu, Chesler:2008hg}. This has given rise to an
extensive literature on holographic thermalization (see,
e.g. \cite{Balasubramanian:2010ce,Das:2010yw, Balasubramanian:2011ur},
for some of the early papers on the subject). This correspondence has
a direct bearing on the issue of universality of thermalization since
a collapse to a black hole state is also typically associated with
loss of most memory of the collapsing matter. In this paper, we will
find that the final equilibrium state is characterized by an infinite
number of thermodynamic parameters (chemical potentials) which retain
a partial memory of the quench protocol\footnote{For a quench from a
  ground state, the final chemical potentials retain a full memory of
  the quench process. When the initial state is different, the final
  chemical potentials retain partial information about the initial
  state and the quench protocol.}; in the holographic dual, this
corresponds to retention of memory by the final black hole of the
collapsing matter.

A significant step in proving thermalization in a closed 2D system was
taken in a recent paper (MSS) \cite{Mandal:2015jla} (similar results
have subsequently appeared in \cite{Cardy:2015xaa}). MSS considered
1+1 dimensional quenches\footnote{Unless otherwise stated, the spatial
  direction will be regarded as non-compact.}, ending with a critical
post-quench Hamiltonian and made the following assumptions:\\ (a) the
post-quench wavefunction is of the generalized Calabrese-Cardy (gCC)
form\footnote{We will define the boundary
  state with an energy cut-off, $ \exp[-\k_2 H]| Bd \rangle$ as the
  Calabrese-Cardy state $|\psi\ran_{CC}$. These states were introduced
  in \cite{Calabrese:2006quench} to describe 2D critical quenches.}
\begin{align}
|\psi \rangle_{gCC}=| \psi(\k_2, \{\k_n\}) \ran
\equiv  \exp[-\k_2 H - \sum_{n>2} \kappa_n W_n]| Bd \rangle
\label{gCC}
\end{align}
where $W_{n}$ are additional conserved charges in the system (the
results are valid even without the additional charges present in the
system).  It was assumed that the charges are obtained from local
currents. Below, for specificity, we will assume that the system is
integrable, with a ${\mathbb W}_\infty$ algebra\footnote{This clearly
  holds for the theory of free scalars and fermions discussed in this
  paper.} and the $W_n$, $n=2,3,...$ ($W_2=H$) are ${\mathbb
  W}_\infty$ charges.  \\ (b) The spectrum of conformal dimensions in
the post-quench critical theory has a gap.  \\ (c) The dimensionless
parameters $\tilde \kappa_n =\kappa_n/\kappa_2^{n-1}, n>2$ are small
and can be treated perturbatively.  \\ (d) The size $l$ of the
interval is small compared to $\kappa_2$.\footnote{The assumptions (c)
  and (d) were made for technical reasons, which can, in principle, be
  obviated in other methods, e.g. if the higher spin deformations
  $\kappa_{n>2}$ can be represented geometrically (like $\kappa_2$
  which is treated as an imaginary time). Assumption (b) appears to be
  more essential. In case of the scalar field model discussed in the
  present work, this condition implies compactifying the range of
  $\phi$ on a circle.}

\gap1
With these assumptions in place, MSS proved that the reduced
density matrix of an interval of size $l$ in the state \eq{gCC}
asymptotes to that in a GGE \footnote{GGE refers to a generalized
  Gibbs ensemble; see, e.g. \cite{Essler:2014qza} for a
  review. Thermalization to a GGE in the context of an integrable CFT
  was anticipated earlier in
  \cite{Calabrese:2012GGE-II,Caputa:2013eka}, and, for more
  general integrable models, in \cite{Barthel:2008GGE, Cramer:2008GGE,
    Rigol:2007, Rigol:2007a, Iucci:2010GGE, Mussardo:2009GGE,
    Calabrese:2011GGE, Calabrese:2012GGE, Calabrese:2012GGE-II,
    Mandal:2013id, Essler:2014GGE}.}, defined by
\begin{align}
\rho_{_{\rm GGE}}= \frac{e^{-\beta H -\sum_{n=3}^\infty 
\mu_n W_n}}{Z}, \kern5pt \beta= 4 \kappa_2, \; \mu_n= 4 \kappa_n, n>2
\label{GGE}
\end{align} 
with a relaxation rate given by\footnote{To be precise the overlap of
  the square-normalized reduced density matrix in the pure state
  \eq{gCC} with that in the mixed state \eq{GGE}, behaves like $1-
  ({\rm const}) e^{-2 \g t}$. See MSS for more details.}
\begin{align}
&\g = \frac{2\pi}{\b} \left[\Delta + \sum_{n=3}^\infty 
\tilde\mu_n  Q_n + O({\tilde\mu_n}^2)\right],
\kern3pt
\tilde \mu_n \equiv \frac{\mu_n}{\b^{n-1}},
\label{gamma}
\end{align}
where $\Delta, Q_n$ are determined by the conformal dimension and other
${\mathbb W}_\infty$ charges of the most relevant operator of the CFT
(by assumption (b) above, $\Delta>0$). A consequence of this result is
that the expectation value of an arbitrary string of local operators,
which can be enclosed in an interval of length $l$, exponentially
thermalizes to its expectation value in the GGE.

One of the motivations of the present work is to extend the proof of
thermalization, without making the assumptions made in MSS, in theories
of free scalars or fermions with a time-dependent mass $m(t)$ quenched
to $m=0$. We allow for nontrivial pre-quench states.

We proceed in two ways: 

\begin{itemize}

\item We consider arbitrary quench protocols $m(t)$ and arbitrary
  squeezed states as pre-quench states (including the ground state)
  and show that the quench leads to a wavefunction of the gCC
  form. This proves the main ansatz of MSS (assumption (a) above). We
  also show that by judiciously choosing the pre-quench states one can
  satisfy the perturbative assumption (c). Thus, for theories
  satisfying (b) and, for intervals satisfying (d), thermalization
  follows from first principles, using the results of MSS.

\item For specific quench protocols, but with arbitrary pre-quench
  states as above, we compute exact time-dependent correlators, and
  explicitly show thermalization of one- and two-point functions, {\it
    without making any of the assumptions of MSS}.\footnote{Of course,
    as we mentioned above, the assumption (a) about the gCC form of
    the wavefunction is in any case true.}

\end{itemize}

One of the technical advances in this paper is the use of non-trivial
pre-quench states, which we take to be squeezed states. The motivation
for considering this class of states is that besides being technically
accessible, these states are experimentally realizable (see,
e.g. \cite{scully1997quantum, squeezed-2011}) and carry non-trivial
quantum entanglement encoded by the squeezing function. 

We list below some salient features of our analysis:

\begin{enumerate}

\item{\bf Memory retention by the equilibrium ensemble:} By using
  inverse scattering methods applied to the above-mentioned auxiliary
  potential scattering, we are able to relate the post-quench
  wavefunction, in particular $\k_n$-parameters of the gCC state, to
  the quench protocol $m(t)$. In fact, if we start with the ground
  state of the pre-quench Hamiltonian, the $\k_n$ parameters
  completely encode $m(t)$, implying that {\it the equilibrium
    ensemble specified by $\mu_n=4\k_n$, carries a precise memory of
    the quench protocol!} In case we start with a squeezed state, the
  equilibrium ensemble remembers a combination of the quench protocol
  and the knowledge of the initial state.

\item{\bf UV/IR mixing (IR sensitivity to irrelevant operators):} As
  already found in MSS, the relaxation rate of various operators
  \eq{gamma}, which govern late time dynamics, depends on all the
  chemical potentials $\mu_n$, equivalently on the $\k_n$. Now from
  \eq{gCC} it is clear that the $\k_n$ represent perturbing a given
  initial state by higher dimensional (irrelevant) operators. Indeed,
  our computation of the exact correlators, shows that for a large
  class of operators, these correlators at long times and large
  distances, are affected by {\it all these} chemical potentials, in
  apparent contradiction to IR universality (this is elaborated in
  Section \ref{sec:non-wilson}). This phenomenon is actually related
  to the memory retention mentioned above.

\item{\bf Holographic correspondence:} Our results show that for a
  given quench protocol, a GGE with a finite number of specified
  chemical potentials can be obtained by taking the pre-quench state
  to be a suitably chosen squeezed state. By using this result and the
  correspondence shown in MSS between thermalization to GGE and
  quasinormal decay to a higher spin black hole, we infer that higher
  spin black holes with an arbitrary set of chemical potentials get
  related to thermalization of {\it squeezed states} in the field
  theory. 

\end{enumerate}

\gap1

\noindent{\bf Outline:}
The outline and organization of the paper is as follows: \\
\gap1 

In Section \ref{sec:scalar} we consider mass quenches in a free scalar
in two dimensions. We relate the dynamics to an equivalent potential
scattering problem, details of which are provided in 
Appendix \ref{app:schro}. We
find that the exact time-dependent wavefunction can be related to a
Bogoliubov transform of the `out' vacuum (the post-quench ground
state). Using this fact we write down the exact form of the scalar
propagator. These results hold for a general mass quench, including
quenches from a massless to a massless theory. We find that the
quenched state is always describable in terms of a gCC state (using an
application of the BCH formula, as described in Appendix
\ref{app:BCH}).  In Section \ref{sec:tanh} we work all of this out for a
specific quench protocol (i.e. specific time dependence of the mass
parameter).  In Section \ref{sec:squeeze} we consider cases where the
pre-quench state is a squeezed state. We show that this gives us a
large class of initial conditions, by tuning which we can prepare a
quench state in the exact form $\exp[-\sum_n \k_n W_n]|D\ran$ which
has a finite number of given $\k_n$ coefficients.  \\

In Section \ref{sec:fermion} we show how to generalize the above
results to fermions.  \\

In Section \ref{sec:GGE} real time Wightman correlators in a GGE are
computed. In Sections \ref{sec:exact-gr} and \ref{sec:exact-mu4} we work out the scalar propagator for the
specific quench protocol of Section \ref{sec:tanh}. This allows us to
compute various exact correlators, starting either from a ground state
or from specific quench states leading to a gCC state with a finite
number of $\k_n$ parameters. We show that these correlators thermalize
exponentially to a GGE; the relaxation rate is found
non-perturbatively, which agrees with \eq{gamma} in the perturbative
regime.\\

In Section \ref{sec:non-wilson} we show that the IR behaviour of exact
correlators is sensitive to all the chemical potentials even though
these represent perturbation by irrelevant operators. We also show
that the equilibrium ensemble remembers the quench protocol. \\

In Section \ref{sec:discussion} we make concluding remarks and mention
some open problems. In Appendices \ref{app:boson} and
\ref{app:fermion} we discuss some notations and general results about
bosonic and fermionic theories. In Appendix \ref{app:sudden-limit} we
elaborate the precise meaning of the sudden limit, taking into account
the UV cut-off of the theory.

\section{\label{sec:scalar}Critical quench of a scalar field: general
strategy}

An important example of quantum quench is provided by free scalar
field theories with time-dependent mass (our notations will closely
follow \cite{Das:2014jna, Das:2014hqa}, which also contain an
extensive reference to the relevant literature).
\begin{align}
S &= -\frac12 \int d^2 x (\del_\mu \phi~ \del^\mu \phi - m^2(t)
\phi^2)\nonumber\\
&= \frac12 \int \frac{dkdt}{2\pi} 
\left(|\dot\phi(k,t)|^2 - (k^2+ m^2(t))
|\phi(k,t)|^2\right), \kern5pt \phi(-k,t)= \phi^*(k,t)
\label{scalar-action}
\end{align}
We will always be working in the thermodynamic limit of infinite
system size. The momentum integrals will be taken to be
$\int^{\infty}_{-\infty} dk/2\pi$, unless otherwise
specified.\footnote{We will sometimes use a large system size $L$, so
  that the momentum integral is replaced by
  $\frac{1}{L}\sum_{n=-\infty}^{\infty}$, such as
  later in this section and in Appendix \ref{app:boson}; however, in
  all these contexts, $L$ will be assumed to be the largest relevant length
  scale. In Appendix \ref{app:sudden-limit} we will use an explicit UV
  cut-off, $|k| < \Lambda$, on the momentum integrals.}

In this section we will consider a mass function $m(t)$ (this is
referred to as a `quench protocol') which decreases from an asymptotic
value $m_0$ in the past to the asymptotic value $m=0$ in the
future. This is called a critical quench since the mass gap vanishes
following the quench.  The generalization to other cases like massless
to massless quench as in Figure \ref{schro-2-fig} is straightforward
and will be touched upon in a later section.
 
Because of translational symmetry, the equations of motion of various
Fourier modes in \eq{scalar-action} get decoupled, where each mode
satisfies a Schr{\"o}dinger-type equation:
\begin{align}
-\ \frac{d^2}{dt^2} \phi(k,t) =(k^2+m^2(t)) \phi(k,t)
\label{eom1}
\end{align}
From the asymptotic behaviour of the potential, it is clear that
\eq{eom1} admits solutions (see, e.g. \cite{birrell1984quantum}) which
behave like plane waves at far past and far future. We will call
$u_{in}(k, t)$ the solution of the wave equation which approaches a
purely positive frequency solution in the past and $u_{out}(k, t)$ the
solution which approaches a purely positive frequency solution in the
future:\footnote{\label{ftnt:sign1}We consider $\exp(-i \omega t)$ and
  $\exp(i \omega t)$ to be future and past directed respectively, with
  energy defined by $i\del/\del t$.}
\begin{align}
u_{in} \xrightarrow{t\to -\infty} \frac{e^{-i \omega_{in} t}}{ \sqrt{2\omega_{in}}},
\quad
u_{out} \xrightarrow{t\to \infty} \frac{e^{-i \omega_{out}t}}{\sqrt{2\omega_{out}}}
\label{def-in-out}
\end{align} 

\begin{figure}[H]
\centering
\includegraphics[scale=.4]{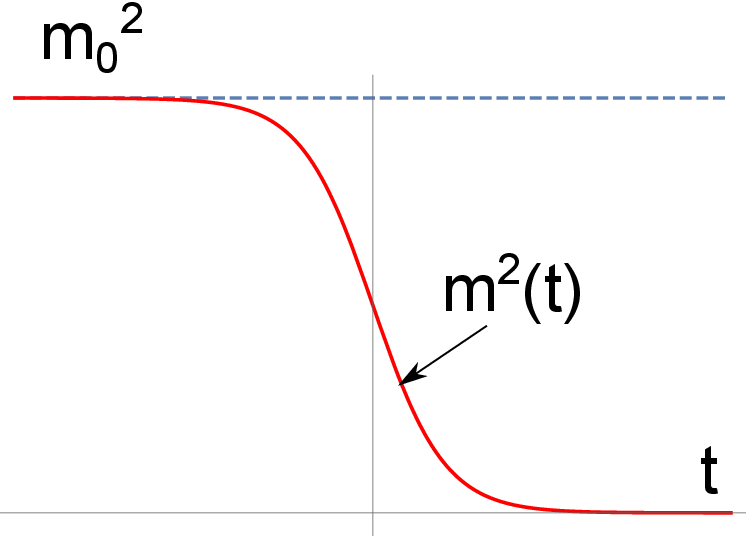}
\caption{Mass quench from $m_0$ to 0.}
\label{schro-fig}
\end{figure}

The question of existence of the exact solutions $u_{in}, u_{out}$ is
essentially identical to the existence of the so-called Jost functions
in the analogous Schr\"odinger problem, which is discussed in detail
in Appendix \ref{app:schro}. The relevant properties of the Jost
functions, and equivalently of the solutions $u_{in}, u_{out}$, can be
proved under appropriate fall-off conditions of the mass-function (to
the asymptotic values) as discussed in the Appendix.

It is easy to see that the pair of solutions $u_{in}(k, t)$ and
$u^*_{in}(-k,t)$ are independent functions (e.g. by computing the
Wronskian in the far past), and hence form a basis of solutions of the
second order differential equation \eq{eom1}. Similar remarks apply
to the pair of solutions $u_{out}, u^*_{out}$. Thus, we can write any
set as a linear combination of the other; e.g.
\begin{align}
u_{in}(k) =  \a(k) u_{out}(k) + \b(k) u^*_{out}(-k),\quad
u_{out}(k) =  \a^*(k) u_{in}(k) - \b(k) u^*_{in}(-k)
\label{uin-uout}
\end{align}
The coefficients $\a, \b$ are called Bogoliubov coefficients (our
conventions here are the same as in \cite{birrell1984quantum}),
which are determined by the mass-function $m^2(t)$.

Indeed, a general solution of the wave equation can be written
as a linear combination of either pair:
\begin{align}
\phi(k,t)= a_{in}(k) u_{in}(k,t) +  a^*_{in}(-k) u^*_{in}(-k,t) 
=a_{out}(k) u_{out}(k,t) +  a^*_{out}(-k) u^*_{out}(-k,t)
\label{in-out-app}
\end{align}
Here we have explicitly put in the reality condition on the scalar
field $\phi(x,t)$ which translates in mixed Fourier space to
$\phi^*(k,t)= \phi(-k,t)$.

Using the two equations above, one can find 
\begin{align}
\label{bogo} a_{in}(k) &= \alpha^*(k) a_{out}(k) - \beta^*(k) a^\dagger_{out}(-k),
\\
\label{inv-bogo}a_{out}(k) &= \alpha(k) a_{in}(k) + \beta^*(k) a^\dagger_{in}(-k),
\end{align}
Upon quantization, the coefficients $a_{in, out}(k)$ are treated as
operators in the Fock space (with $a^*_{in, out}(k)$ rewritten as
$a^\dagger_{in, out}(k) $). The specific normalizations of the
$u$-functions in \eq{def-in-out} imply that the Bogoliubov coefficients
satisfy the following constraints 
\begin{align}
|\a|^2 - |\b|^2=1
\label{asq-bsq}
\end{align}
The derivation of the above relations can be found in standard
textbooks, e.g. \cite{birrell1984quantum}, or from the 
analogous Schr\"odinger problem as in Appendix \ref{app:schro}.

\subsection{\label{sec:ground-gCC}General proof of the gCC ansatz 
\cite{Mandal:2015jla} for the ground state}

The two sets of oscillators $a_{in}(k)$ and $a_{out}(k)$ define two
distinct vacua $| 0, in\ran$ and $|0, out\ran$, defined by $a_{in}(k)|
0, in\ran=0$ and $a_{out}(k)| 0, out \ran=0$. Let us assume that we
start from the ground state of the original massive theory,
i.e. $|0,in\ran$.

Using \eq{bogo}, we can express the in-vacuum in terms of the
out-vacua as follows\footnote{This is proved by simply checking that
  the right hand side is annihilated by $\alpha^*(k) a_{out}(k) -
  \beta^*(k) a^\dagger_{out}(-k)$. Here $\sum_k$ is defined as the sum
  over discretized values of $k$, as elaborated in Appendix
  \ref{app:boson}.}
\begin{align}
| 0, in \ran & = \exp[\frac12 \sum_k \gamma(k) a^\dagger_{out}(k)
  a^\dagger_{out}(-k)] | 0, out \ran, 
\label{in-out-vac}
\end{align}
where
\begin{align} 
\gamma(k) =  \beta^*(k)/\alpha^*(k)
\label{gamma-k}
\end{align} 

With the above ingredients in place, it's a simple exercise, using the
Baker-Campbell-Hausdorff formula (see Appendix \ref{app:BCH}), to show
that the in-vacuum can be written in the following form\footnote{\label{ftnt:diptarka}This result was independently found
  some time ago, for the quench protocol discussed in Section
  \ref{sec:tanh}, in \cite{Diptarka:BCH}. We thank Sumit Das for
  sharing these results with us.}
\begin{align}
|0, in \ran&= \exp[\frac12 \sum_k \g(k) a_{out}^\dagger
 (k)a_{out}^\dagger (-k)] |0, out\ran = \exp[-\sum_k \kappa(k)
a_{out}^\dagger (k)a_{out}(k)] | D \ran,\;
\label{gen-gCC}\\
\kappa(k) &=- \frac12 \log(- \g(k))
\label{BCH}
\end{align}
where $|D\ran$ is a Dirichlet boundary state \eq{dirichlet}, defined
in terms of the `out' Fock space: 
\begin{align}
|D\ran = \exp\left[-\frac12 \sum_{k} a^\dagger_{out}(k)
  a^\dagger_{out}(-k)\right]|0, out \ran.
\label{dirichlet-out}
\end{align}
We will identify the right hand side of \eq{gen-gCC} as a gCC
(generalized Calabrese-Cardy) state ({\it cf.}  \eq{gCC}) in which the
charges are expressed as a momentum integral and the boundary
state is identified with a Dirichlet state:
\begin{align}
|0, in \ran &= |\psi\ran_{gCC}
\nonumber\\
|\psi\ran_{gCC} &=  \exp[-\sum_k \kappa(k) N(k)
] | D \ran, \quad N(k)=  a_{out}^\dagger (k)a_{out}(k)
\label{gen-gCC-def}
\end{align}
We will explain below (see under `Interpretation') that this
shows our desired result, namely quantum quench in the free
scalar theory with mass $\to$ zero leads to a gCC state.

\paragraph{Expanded form of the gCC state:}

In Appendix \ref{app:implications} we have shown, by borrowing results
from the analogous Schrodinger problem, that $\g(k)$ admits a
small-momentum expansion of the form
\begin{align}
\g(k)=-1 + \g_1 | k| + \g_2 |k|^2 +\g_3 |k|^3 + ..., 
\label{gamma-expansion-text}
\end{align}
Using this power series expansion, and the
expression for $\k(k)$ in \eq{BCH}, we can expand $\k(k)$ also in a
power series around $k=0$, as follows:
\begin{align}
\kappa(k) &= \kappa_2 |k| + \kappa_3 |k|^2 + \kappa_4 |k|^3 - ...,
\nonumber\\ \k_2 &= \frac{\g_1}{2}, \k_3= \frac{1}{4} \left(\g_1^2+ 2
\g_2\right), \k_4= \frac{1}{6} \left(\g_1^3+ 3 \g_1 \g_2 +
3\g_3\right), ...
\label{kappa-expansion}
\end{align}
The radius of convergence of the above expansion is determined by the
singularity of $\gamma(k)$ in the complex $k$ plane which is nearest
to the origin. The location of this is determined by the parameters
with mass dimension that characterize the mass function $m^2(t)$. In
case of a single scale $m_0$ we will find that the radius of
convergence equals $m_0$ (see remarks below
\eq{gamma-expansion-tanh}). The expansion above is therefore convergent
for small $k$ (smaller than the parameters of mass dimension
characterizing the massive phase).

Below we will find explicit examples of this power series for specific
quench protocols $m(t)$ which interpolate from $m_0$ to $m=0$. For
quenches involving a single real scalar field, we will find that the
above expansion \eq{kappa-expansion} has only odd powers of
$|k|$,\footnote{This is consistent with the fact that a real scalar
  field provides a representation of the $W_\infty$ algebra
  \cite{Bakas:1990ry} where the odd $W_n$'s vanish.  See below.} and,
explicitly $\kappa_2>0$.\footnote{For massless$\to$massless quench,
  $\k_2$ turns out to be purely imaginary (see Section
  \ref{sec:critical}).}  Substituting such an expansion for
$\kappa(k)$ in \eq{gen-gCC-def}, we find 
\begin{align}
|0, in \ran =  \exp[-\kappa_2 H -\sum_{n=2}^\infty \kappa_{2n}W_{2n}] 
| D \ran   
\label{ground-gCC}
\end{align}
where $W_{2n}, n=1,2,...,$($W_2=H$) are the even $W_\infty$ charges
\cite{Bakas:1990ry} of the final massless scalar field theory, which
we define here as follows\footnote{The normalization convention here
  for the $W$-charges differs from that of
  \cite{Bakas:1990ry}.}\fnsep\footnote{\label{ftnt:w-inf}If the
  time-dependence of the Hamiltonian stops after a finite time, the
  post-quench Hamiltonian coincides with the $W_2$ charge, and the
  other $W_{2n}$ charges also represent conserved charges of the
  post-quench evolution.}
\begin{align}
H \equiv W_2 = \sum_k  |k| a_{out}^\dagger(k)a_{out}(k), \;
W_{2n} =  \sum_k  |k|^{2n-1} a_{out}^\dagger(k)a_{out}(k),\, n=2,3,...
\label{charges}
\end{align}
As discussed above, the expansion \eq{kappa-expansion} has a finite
radius of convergence; putting such a power series inside the
$k$-integral in \eq{gen-gCC-def} appears, {\it a priori}, to be
problematic. However, the terms in the resulting series, as in
\eq{ground-gCC}, involve $\kappa_{2n} W_{2n}$ where $W_{2n}$ are
operators and not numbers. It is important to consider correlators or
expectation values and check the resulting series for convergence. In
practical calculations, such as the calculation of correlators in
later sections, we will more often use \eq{gen-gCC-def} directly than
the form \eq{ground-gCC} (similar statements can be made in the
context of the GGE ensemble, which can be defined in the sense of a
momentum-integral; for example of such calculations, see \eq{gen-GGE}
and \eq{del-del-comparison}).  Furthermore, one can clearly construct
overlap of \eq{ground-gCC} with states which have support only over a
finite range of momenta less than $m_0$, in which case there is no
problem of convergence of the series in \eq{ground-gCC}. Additionally,
we will find in later sections that one can explicitly construct
examples of quenches (starting from squeezed states) where the series
$\sum_n \kappa_{2n} W_{2n}$ terminates after a finite number of terms;
in that case also, the series clearly makes sense.  Henceforth, we
will consider the series expression in \eq{ground-gCC} with these
qualifiers in mind.\footnote{We thank the referee for raising this
  point.}

\paragraph{Interpretation:}

\noindent To interpret the result \eq{gen-gCC-def} or \eq{ground-gCC},
let us first assume a sudden quench.

In this case, the state we obtain immediately after the quench is
still the initial ground state $|0, in \ran$. {\it This proves that
  the post-quench wavefunction is indeed of the gCC form
  \eq{gen-gCC-def} or \eq{gCC} as claimed in the introduction.}

\noindent Note that the values of the charges are given by
\begin{align}
\kern-20pt\lan W_{2l} \ran =  \sum_k  |k|^{2l-1} \lan N(k) \ran,\, l=1, 2,...,
\hbox{where}~ \lan N(k) \ran 
\equiv  \lan 0_{in}| a_{out}^\dagger(k)a_{out}(k) | 0_{in} \ran
= | \b(k) |^2  
\label{charge-values}
\end{align} 
The last step famously follows by expressing the `out'-oscillators
in terms of the `in'-oscillators using \eq{bogo}.

In case the quench is not sudden, we proceed as follows. Let us
consider the Wightman function
\begin{align}
\lan 0, in |  O_1(x_1, t_1) O_2(x_2, t_2) ... O_n(x_n, t_n) |0, in \ran
\label{wightmann}
\end{align}
where $O_i$ are some operators built of the field $\phi(x,t)$ and its
derivatives. The one- and two-point functions considered in Section
\ref{sec:corr} are examples of this. 

Let us first assume that the quench takes place for a finite period of
time, up to a time $t_0$. In case the time instants $t_i$ are all in
the post-quench period, we can express all operators in terms of the
{\it out} Heisenberg oscillators $a_{out}, a^\dagger_{out}$ with
simple exponential time-dependence. If we express the state $| 0, in
\ran$ as in \eq{ground-gCC}, the result of this exercise will be a
calculation of {\it out} Heisenberg oscillators as if the post-quench
state is of the gCC form \eq{gCC}. This is the viewpoint adopted
in standard textbooks of quantum field theory in curved space time
(e.g. \cite{birrell1984quantum}).

In case the quench stops only asymptotically, but sufficiently fast,
the above statement goes through for time instants $t_i$ which are
late enough.\footnote{In the explicit examples considered in the paper, the
mass function has an exponential tail, of the form $e^{-\rho t}$;
thus the gCC ansatz works to an exponential accuracy, up to 
terms $O(e^{-\rho t_i})$ which can be made arbitrarily small by considering
time instants $t_i \gg 1/\rho$. \label{ftnt:asymptote}}. 

~\\ {\bf Conclusion:} {\it Thus, we find that the post-quench
  wavefunction, starting from the ground state of the original
  Hamiltonian, under a quantum quench to zero mass, is represented in
  the generalized Calabrese-Cardy (gCC) form, as predicted in
  \cite{Mandal:2015jla}.}

We will find below that the above conclusion also holds when we start
from more general states in the initial massive theory.


\subsection{Thermalization to GGE\label{sec:eqm}}

In this subsection, we briefly recall results from MSS
\cite{Mandal:2015jla} on ``subsystem thermalization'' and their
implications in the present case.

MSS considered the time evolution of a subsystem $A$ (which remains
finite in the thermodynamic limit) in a quantum quench.  For
post-quench gCC-type states \eq{gCC} and for a perturbative domain in
the $\k_n$ parameters, it was shown that the reduced density matrix
(RDM) of the region $A$ (obtained after tracing out $A^c$, the
complement of $A$) asymptotically approaches the RDM of $A$ in a GGE
\eq{GGE}:
\begin{align}
\text{Tr}_{A^c}\left[\text{e}^{-iHt}| \psi(\k_2, \{\k_n\}) \ran\lan\psi(\k_2, \{\k_n\})|\text{e}^{iHt}\right] 
\xrightarrow{t\to\infty} \text{Tr}_{A^c}\left[\rho_{_{\rm GGE}}(\b,\{\mu_n\})
\right]\quad \b=4\k_2, \mu_n =4\k_n
\label{thermalization}
\end{align}
More explicitly, 
\begin{align}
\text{Tr}_{A^c}\left[\text{e}^{-iHt}e^{-\sum_k \k(k) \hat N(k)} | D\ran\lan D|e^{-\sum_k \k(k) \hat N(k)}\text{e}^{iHt}\right]
\xrightarrow{t\to\infty} \; \text{Tr}_{A^c} 
\left[\f1Z e^{-\sum_k \mu(k) \hat N(k)}\right]
\label{thermalization-mode}
\end{align} 
The above statement is called {\it subsystem thermalization}. Below
we will calculate time-dependent Wightman functions to explicitly
verify this.

The energy and $W$-charges (as well as the number operator) are
conserved in the post-quench CFT dynamics. The above statement
implies that
\begin{align}
\vev{H}_{_{\rm gCC}} = \vev{H}_{_{\rm GGE}}, \; 
\vev{W_n}_{_{\rm  gCC}} = \vev{W_n}_{_{\rm GGE}},\;
\vev{N(k)}_{_{\rm gCC}} = \vev{N(k)}_{_{\rm GGE}}
\label{conserved}
\end{align}
Thus, the charges \eq{charge-values} measured for the post-quench
(gCC) 
state are the same as those of the GGE. In particular, using
\eq{asq-bsq}, \eq{gamma-k} and \eq{BCH}, it is easy to see that
\begin{align}
\vev{N(k)} = |\b(k)|^2 = \f{|\g(k)|^2}{1- |\g(k)|^2} = 
\f1{e^{4\k(k)}-1}= \f1{e^{\mu(k)}-1}, \; \mu(k) \equiv 4\k(k).
\label{rigol-chem-pot}
\end{align}
The last expression gives a Bose distribution for each $k$, as
appropriate for a GGE  \cite{Rigol:2007}.

\subsection{\label{sec:prop}The propagator}

Using the defining property of the in-vacuum $|0, in \ran$, and the
mode expansion of $\phi(x,t)$ in terms of the in-modes, it is easy to
derive the following basic two-point function
\begin{align}
&\langle 0,in| \phi(x_1,t_1) \phi(x_2,t_2) |0,in\rangle
=\int \frac{dk}{2\pi}\;  u_{in}(k,t_1)u^*_{in}(k,t_2)\ e^{ik(x_1-x_2)}
\nonumber\\
=&\int \frac{dk}{2\pi}\ \left[|\alpha(k)|^2 u_{out}(k,t_1)u^*_{out}(k,t_2)+
\alpha(k) \beta^*(k)u_{out}(k,t_1)u_{out}(-k,t_2) \right.\nonumber\\
& \left.+ \alpha^*(k)\beta(k) u^*_{out}(-k,t_1)u^*_{out}(k,t_2)+
|\beta(k)|^2u^*_{out}(-k,t_1)u_{out}(-k,t_2)\right]e^{ik(x_1-x_2)}
\label{phi-phi}
\end{align}
In the second step we have used the relation \eq{uin-uout} between the
`in' and `out' modes. Interestingly the particular combination
of Bogoliubov coefficients appearing above can be written entirely in terms
of $\g(k)$ (by using \eq{gamma-k} and \eq{asq-bsq}):
\begin{align}
&|\alpha(k)|^2= \frac1{1- | \g(k)|^2},\; 
|\b(k)|^2= \frac{| \g(k)|^2}{1- | \g(k)|^2},\nonumber\\
&\alpha(k)\b^*(k)=  \frac{ \g(k)}{1- | \g(k)|^2},\; 
\alpha^*(k)\b(k)=  \frac{\g^*(k)}{1- | \g(k)|^2}
\label{a-b-ident}
\end{align}
The propagator \eq{phi-phi} has recently appeared in
\cite{Das:2015jka} where it is used to study the relation between
smooth fast quenches and instantaneous quenches. Related expressions,
in a somewhat different form, have appeared in
\cite{Sotiriadis:2010si}.\\ In Section \ref{sec:exact-gr} we will
determine this propagator exactly for a specific quench protocol.

\subsection{\label{sec:tanh}A specific quench protocol}

We will now work out some of the above ideas for the specific mass
function
\begin{align}
m^2(t)= m_0^2 (1- \tanh(\rho t))/2
\label{tanh}
\end{align}
The equation of motion \eq{eom1} with the above mass profile can be exactly solved
(see, e.g. \cite{birrell1984quantum}, Chapter 3, where this model
appears in a simple model of cosmological particle creation). Using
this fact, we can find the following explicit solutions for
$u_{in}(k,t)$ and $u_{out}(k,t)$:
\begin{align}
u_{in}(k,t)&= \frac{e^{-i\omega_{in}t}}{\sqrt{2 \omega_{in}}} \; _2F_1\left(\frac{i\omega_-}{\rho },-\frac{i\omega_+}{\rho };1-\frac{i\omega_{in}}{\rho };-e^{2\rho t}\right)
\\
u_{out}(k,t)& =\frac{e^{-i\omega_{out}t}}{\sqrt{2 \omega_{out}}} \; _2F_1\left(\frac{i\omega_-}{\rho},
\frac{i\omega_+}{\rho};\frac{i\omega_{out}}{\rho}+1;-e^{-2\rho t}\right)
\end{align}
where $_2F_1$ is a hypergeometric function and
\begin{align}
\omega_{in}=\sqrt{k^2+m_0^2},\quad \omega_{out}=|k|, 
\quad \omega_{\pm}=\frac{1}{2}(\omega_{out}\pm\omega_{in})\nonumber\
\end{align}
Using \eq{uin-uout} and properties of hypergeometric functions \cite{Abramowitz} for
large arguments, we find the following Bogoliubov coefficients
\begin{align}
\alpha(k) &=\sqrt{\frac{\omega_{out}}{\omega_{in}}}\; \frac{\Gamma \left(-\frac{i \om_{out}}{\rho }\right) \Gamma \left(1-\frac{i \om_{in}}{\rho }\right)}{\Gamma \left(-\frac{i \om_+}{2 \rho }\right) 
\Gamma \left(1-\frac{i \om_+}{2 \rho }\right)}, \quad
\beta(k) =\sqrt{\frac{\omega_{out}}{\omega_{in}}}\;  \frac{\Gamma 
\left(\frac{i \om_{out}}{\rho }\right) \Gamma \left(1-\frac{i \om_{in}}{\rho }\right)}{\Gamma \left(\frac{i \om_-}{2 \rho }\right)
\Gamma \left(1+ \frac{i \om_-}{2 \rho }\right)}\nonumber\
\end{align}
which gives
\begin{align}
\g=\f{\b^*}{\a^*}=& -1
+ 2\left(\f{k}{m_0}\right) \left(-1 + \frac{2 i m_0}{\rho} 
\left(\gamma +\psi ^{(0)}(-\frac{i m}{2 \rho })\right)\right)
\nonumber\\
&+
2\left(\frac{k}{m_0}\right)^2 \left(\f{m_0}{\rho} [\gamma
   +\psi ^{(0)}(-\frac{i m}{2 \rho })]+i \right)^2+
O\left(\f{k}{m_0}\right)^3
\label{gamma-expansion-tanh}
\end{align}
The power series expansion here is consistent with the general form
\eq{gamma-expansion-text}. Indeed, from the explicit expression of the
Bogoliubov coefficients, $\g(k)$ can be seen to be analytic in the
complex $k$ plane near the origin, with the nearest singularity given
by $k= \pm i m_0$. 
 
Applying the general method of Section \ref{sec:ground-gCC} to this case,
we find that the ground state is explicitly of the gCC form
\eq{ground-gCC},
\[
|0, in \ran =  \exp[-\kappa_2 H -\sum_{n=2}^\infty \kappa_{2n}W_{2n}] 
| D \ran  
\]
where the $\kappa_n$'s are found by using \eq{BCH}. In an expansion in
$m_0/\rho$ (to be interpreted in the sense of Appendix
\ref{app:sudden-limit}), these coefficients read as follows
\begin{align}
&\kappa_2 =\frac{1}{m_0}\left(1 + \frac{\pi^2}{12}\left(\f{m_0}{\rho}
\right)^2 - i\f{\zeta(3)}{4} \left(\frac{m_0}{\rho}\right)^3 -\frac{\pi^4}{720m_0}\left(\f{m_0}{\rho}
\right)^4+
O\left(\frac{m_0}{\rho}\right)^5\right),\nonumber\\ 
& \kappa_4=
\f1{m_0^3}\left(-\frac{1}{6} + \f{\pi^2}{24}\left(\f{m_0}{\rho}
\right)^2 -\f{\pi^4}{288}\left(\f{m_0}{\rho}
\right)^4+ O\left(\frac{m_0}{\rho}\right)^5\right), \, ...
\label{kappa-s}
\end{align}
The coefficients $\kappa_n$ \eq{kappa-s} are functions of both the
scales $m_0$ and $\rho$; note that the knowledge of these coefficients (in
this case the first two, $\k_2$ and $\k_4$) encode the quench protocol
\eq{tanh} completely. Since the $\k_n$'s are related in a one-to-one
fashion to equilibrium chemical potentials $\mu_n = 4\k_n$ \eq{GGE},
it follows that from the equilibrium state one can retrieve the quench
history (see Section \ref{sec:memory} for more details). ~\\ For later
reference, the ``out''-number operator \eq{charge-values} turns out to
be
\begin{align}
\lan N(k) \ran =
\text{csch}\left(\frac{\pi  k}{\rho }\right) \sinh ^2\left(\frac{\pi 
   \left(k-\sqrt{k^2+m_0^2}\right)}{2 \rho }\right) \text{csch}\left(\frac{\pi 
   \sqrt{k^2+m_0^2}}{\rho }\right)
\label{number-tanh}
\end{align}
~\\ {\it We thus explicitly verify here that the post-quench state is
  of the form \eq{gCC}.}\footnote{\label{post-quench}In the sense of
  the comments following \eq{wightmann}.}

\subsubsection{\label{sec:sudden}Sudden limit}

We will be especially interested in the sudden limit
\begin{align}
\rho \to \infty
\label{naive-sudden}
\end{align} 
which gives the simple quench protocol
\begin{align}
m^2(t)= m_0^2 \Theta(-t),
\label{tanh-sudden}
\end{align}
where $\Theta(t)$ is the Heaviside step function. In this limit the
Bogoliubov coefficients become\footnote{These can be compared with the
  corresponding quantities of an analogous Schr\"odinger problem is
  discussed in Section \ref{exPots}, example 1.}
\begin{align}
 \alpha(k)&=\frac{1}{2}\ \frac{|k|+\om_{in}}{\sqrt{|k|\om_{in}}},
\; \beta(k)=\frac{1}{2}\ \frac{|k|-\om_{in}}{\sqrt{|k|\om_{in}}},
 \label{ab-sudden}
\end{align}
and the in- and out- waves reduce to
\begin{align}
u_{in}(k,t)= \frac{e^{-i\omega_{in}t}}{\sqrt{2 \omega_{in}}},\;
u_{out}(k,t) =\frac{e^{-i\omega_{out}t}}{\sqrt{2 \omega_{out}}} 
\label{wave-sudden}
\end{align}
The $\k_n$ coefficients in the sudden limit are given by taking the
$\rho\to\infty$ limit of \eq{kappa-s}:
\begin{align}
\kappa_2 =\frac{1}{m_0},\,  
\kappa_4= -\f1{6m_0^3}, \, ...
\label{kappa-sudden}
\end{align}
Thus, 
\begin{align}
|0, in \ran =  \exp\left[-\frac{H}{m_0}+ \frac{W_4}{6 m_0^3} + \cdots\right]
| D \ran  
\label{gCC-sudden}
\end{align}
which is a gCC state.\footnote{\label{ftnt:norm}One might be alarmed
  by the positive sign of the $W_4$-coefficient in this state. This
  would mean that if all the higher $\k_{n>3}$ were absent, $\k(k)$
  would have grown as $+k^3$, hence implying a divergent norm of the
  gCC state $e^{-\sum_k \k(k) N(k)}|D \ran$. However, such
  catastrophes are avoided by higher $\k_n$ coefficients, as they
  must, since the gCC state is equal, as a Heisenberg state, to the
  initial ground state, which has a finite norm. We will have more to
  say in Appendix \ref{app:sudden-limit} on other possible divergences
  associated with the sudden limit.}  In the sudden limit, the number
operator \eq{number-tanh} becomes
\begin{align}
\lan N(k) \ran =   \frac{\left(\sqrt{k^2+m_0^2}- |k|\right)^2}{
4 \sqrt{k^2+m_0^2} |k|}
\label{number-sudden}
\end{align}
Strictly speaking, the sudden limit, as defined by \eq{naive-sudden},
is somewhat naive, and needs to be refined, keeping the UV cut-off of
the theory in mind. A precise and careful version is presented in
Appendix \ref{app:sudden-limit}. One result of that analysis is that
the naive sudden limit \eq{naive-sudden} gives the correct results for
most considerations in this paper (Appendix \ref{app:sudden-limit}
describes some cases where more care is needed).

\subsection{\label{sec:critical}Quenching from critical to critical}
In this subsection, we will consider a quantum quench for the scalar
field where both the initial and final masses vanish (i.e. a quench
from a critical Hamiltonian to a critical Hamiltonian).
 
\begin{figure}[H]
\centering
\includegraphics[scale=.4]{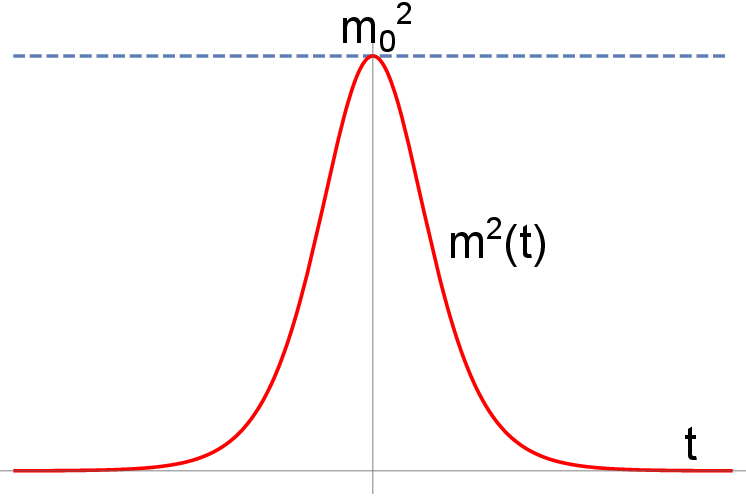}
\caption{\footnotesize A mass-profile describing quantum quench from a critical
  Hamiltonian back to the critical Hamiltonian.  Here $m^2(t)
  \xrightarrow{t\to \pm \infty} 0$.}
\label{schro-2-fig}
\end{figure}

\noindent A typical mass function which follows this property is
\cite{Das:2014hqa}:
\begin{equation}
m^2(t)=m_0^2\ \sech^2(\rho t).
\end{equation}
Using the coordinate transformation $y=e^{2\rho t}$. The equation
of motion, analogous to \eq{eom1}, becomes
\begin{equation}
\phi''(k,y) + \frac{\phi'(k,y)}{y} + \left(\frac{k^2}{4\rho^2y} 
+ \frac{m_0^2}{\rho^2(1 + y)^2}\right) \phi(k,y) = 0
\end{equation}
With $\alpha=1/2 + \frac{1}{\rho}\sqrt{4 m_0^2 + \rho^2}$, 
this equation can be solved to give
\begin{align}
u(k,t)&=e^{-ikt} (1 + e^{2\rho t})^{\alpha}\left[C_1\ e^{2ikt}\ {}_2F_1\left(
\alpha , \frac{ik}{\rho} + \alpha, 1 + \frac{ik}{\rho}, -e^{2\rho t}\right) \right. \nonumber \\
&+ \left. C_2\, {}_2F_1\left(\alpha , -\frac{ik}{\rho} +\alpha , 1 - \frac{ik}{\rho}, -e^{2\rho t}\right)\right]
\end{align}
$C_1=1$ and $C_2=0$ gives the incoming solution $u_{in}(k)$ which
satisfies the property \eq{def-in-out}. On taking the $t\to +\infty$
limit of $u_{in}(k)$ we can express $u_{in}(k)$ in the form $\alpha(k)
u_{out}(k) + \beta(k) u^*_{out}(k)$ as in \eq{uin-uout}, where
\begin{align}
\alpha(k)&=\frac{\Gamma \left(\frac{i k}{\rho} +1\right) \Gamma \left(\frac{i k}{\rho}\right)}{\Gamma \left(\frac{i k}{\rho}-\alpha +1\right)
   \Gamma \left(\frac{i k}{\rho} +\alpha \right)}\\
\beta(k)&=i \sin (\pi  \alpha ) \text{cosech}\left(\frac{\pi  k}{\rho}\right)
\end{align}
Using \eq{gamma-k} and \eq{BCH}, we can express the in-vacuum in a gCC
form \eq{gCC} with
\begin{equation}
\kappa(k)=\frac{i k \rho}{2 m_0^2}-\frac{k^2 \rho^2}{4 m_0^4}-
\frac{i k^3 \rho^3}{6  m_0^6}+\frac{k^4 \rho^4}{8 m_0^8}+
\frac{i k^5 \rho^5}{10 m_0^{10}} + \dots,
\end{equation}
which leads to 
\[
\kappa_2 = \frac{i\rho}{2m_0^2}, \kappa_3=\frac{-\rho^2}{4 m_0^4}.
\]
Note that $\kappa_2$ is imaginary. By contrast, in a massive $\to$
massless quench, $\k_2$ is real and positive (see
e.g. \eq{kappa-sudden}), and is identified with $\beta/4$ where $\b$
is the inverse temperature of the associated thermal state. With
imaginary $\k_2$, such an identification is clearly problematic. We
will find in the next subsection that starting with an appropriate
squeezed state in a massless theory and using the above quench
protocol, one can manufacture a CC state with positive $\k_2$.

\subsection{\label{sec:squeeze}Quenching squeezed states}

In this subsection, we will show that gCC states can result even from
excited states of the initial Hamiltonian. In particular, we will find
that specific choice of such initial states can lead to CC states
where $\k_2 \ne 0$, but all other $\k_n=0$.

Suppose, instead of the ground state we start with an arbitrary squeezed
state\footnote{These states have importance in diverse contexts
  \cite{scully1997quantum, perelomov1986generalized} including quantum
  entanglement \cite{squeezed-2011}. Time-development of these states
  can address the issue of dynamical evolution of quantum
  entanglement, among other things.}  of the pre-quench
Hamiltonian:\footnote{\label{ftnt:norm-sq}We assume that the norm of
  the squeezed state is finite, which is ensured by the finiteness of
  the integral $\int dk/(2\pi) \log(1- | f(k)|^2)$.}
\begin{align}
| \psi, in \ran = |f\rangle
\equiv \exp\left[\frac12 \sum_k f(k) a_{in}^\dagger(k) 
a_{in}^\dagger(-k)\right]| 0,in \rangle
\label{squeezed}
\end{align}
This is clearly a Bogoliubov transformation of $|0, in\ran$. To see
this, note that $|f\ran$ is annihilated by $a_{in}(k) - f(k)
a_{in}^\dagger(-k)$,
\begin{eqnarray}
0&=&\left[a_{in}(k)-f(k)a^\dagger_{in}(-k)\right]|f\rangle\nonumber\\
&=&\left[\alpha^*(k)a_{out}(k)-\beta^*(k)a^\dagger_{out}(-k)-f(k)\left\{\alpha(k)a^\dagger_{out}(-k)-\beta(k)a_{out}(k)\right\}\right]|f\rangle\nonumber\\
&=&\left[\left\{\alpha^*(k)+f(k)\beta(k)\right\}a_{out}(k)-\left\{\beta^*(k)+f(k)\alpha(k)\right\}a^\dagger_{out}(-k)\right]|f\rangle\
\end{eqnarray}

Thus, it follows that  the squeezed pre-quench
state is also expressible as a generalized CC state
\begin{align}
& |f\rangle= \exp\left[\frac12 \sum_k \g_{_{\rm
      eff}}(k) a_{out}^\dagger(k) a_{out}^\dagger(-k)\right]| 0,out\rangle 
\label{pre-squeeze}
\end{align}
where the effective $\g_{_{\rm eff}}(k)$ is
\begin{align}
\g_{_{\rm eff}}(k)
=\frac{\beta^*(k) + f(k)\alpha(k)}{\alpha^*(k)+ f(k) \beta(k)}
= e^{i \delta(k)} \f{\g(k)  + f(k) e^{i \delta(k)}}{1 +  e^{i \delta(k)} f(k) \g^*(k)},
\;   e^{i \delta(k)} = \f{\a(k)}{\a^*(k)}
\label{gamma-eff}
\end{align}
Using the result \eq{pre-squeeze} and the method leading to 
\eq{BCH}, we can again show 
\begin{align}
&| f\ran = \exp[\sum_k -\kappa_{_{\rm eff}}(k)
  a_{out}^\dagger (k)a_{out}(-k)] | B \ran,\;
\nonumber\\
&\kappa_{_{\rm eff}}(k) \equiv - \frac12 \log \left(\frac{\g_{_{\rm eff}}(k)}{\g_0}\right)
\label{kappa-eff}
\end{align}
where, as defined in \ref{bdry-state}, for $\g_0=-1$, $|B\rangle$ is the Dirichlet state $|D\rangle$ and for $\g_0=1$, $|B\rangle$ is the Neumann state $|N\rangle$. Using the Taylor expansion
\eq{gamma-expansion-text}, and assuming analyticity of $\delta(k)$ at
$k=0$, we can easily show that $\k_{\rm eff}(k)$ has an expansion of the
form \eq{kappa-expansion} (with the possible addition of a constant
term if $\delta(0)\ne 0$; this will have the interpretation of
adding a chemical potential corresponding to the total number
operator).

\paragraph{\bf Explicit Examples:}

Let us fix the quench protocol to be the sudden limit of the `tanh'
function \eq{tanh-sudden}.  We will determine the $\k_{\rm eff}$
explicitly by using \eq{kappa-eff} and the expressions for the
Bogoliubov coefficients \eq{ab-sudden}.

\begin{itemize}

\item{\bf Gaussian:} For a Gaussian squeezing function with variance
proportional to $m_0^2$, ie. $f(k)=\exp[-k^2/(a^2 m_0^2)]$, we get
\begin{align}
\kappa_{\rm eff}(k)= \frac{|k|}{a^2 m_0}+\frac{\left(6 a^4+1\right)
  |k|^3}{12 a^6 {m_0}^3}-\frac{\left(30 a^8-10 a^4-3\right) |k|^5}{240
  a^{10} {m_0}^5}+ O(|k|^7)
\end{align}
with Neumann boundary state $|B\rangle=|N\rangle$.

\item{\bf Preparing CC states and gCC states with specified
  parameters:} It is clear from \eq{kappa-eff} that given specific
  Bogoliubov coefficients, e.g. \eq{ab-sudden}, we can obtain any
  desired expression for $\k_{\rm eff}(k)$ by tailoring the choice of the
  squeezing function $f(k)$. Thus, e.g.
\begin{align}
f(k)= f_4(k)=
1-\frac{2 |k|}{\sqrt{k^2+{m_0}^2} 
\tanh \left(\k_{2,0}  |k| + \k_{4,0} |k|^3 \right)+|k|}
\label{tailored}
\end{align}
yields a function $\k_{\rm eff}(k)= \k_{2,0} |k| + \k_{4,0} |k|^3$ with
specified parameters $\k_2=\k_{2,0}$,\ $\k_4= \k_{4,0}$ and $\k_n=0$
for $n>4$. This identifies the squeezed state with a gCC state with
these $\k$-parameters:\footnote{Note that we choose here $\k_{2,0}$,
  $\k_{4,0}$ to be positive to ensure that the gCC state is of finite
  norm; see footnote \ref{ftnt:norm-sq}.}
\begin{align}
| \psi, in \ran= | f_4\ran = \exp[-\left(\k_{2,0} H + \k_{4,0} W_4
  \right)| D\ran
\label{tailored-gCC}
\end{align}
In the next subsection, we will specialize to $\k_{4,0}=0$ in
\eq{tailored} to prepare a CC state. Note that the squeezing
functions are localized in $k$ which ensures the normalizability
of the squeezed state.

\end{itemize}

\subsubsection{\label{sec:sq-prop}The propagator in a squeezed state}

The propagator in a squeezed state $| \psi, in\ran = | f\ran$ is
obtained by replacing $\a \to \a_{\rm eff}$, $\b\to \b_{\rm eff}$ in
\eq{phi-phi}:
\begin{align}
&\langle \psi,in|\phi(x_1,t_1)\phi(x_2,t_2)|\psi,in\rangle
\nonumber\\
=&\int \frac{dk}{2\pi}\ \left[|\alpha_{\rm eff}(k)|^2 u_{out}(k,t_1)u^*_{out}(k,t_2)+
\alpha_{\rm eff}(k) \beta_{\rm eff}^*(k)u_{out}(k,t_1)u_{out}(-k,t_2) \right.\nonumber\\
& \left.+ \alpha_{\rm eff}^*(k)\beta_{\rm eff}(k) u^*_{out}(-k,t_1)u^*_{out}(k,t_2)+
|\beta_{\rm eff}(k)|^2u^*_{out}(-k,t_1)u_{out}(-k,t_2)\right]e^{ik(x_1-x_2)}
\label{phi-phi-sq}
\end{align}
Here the bilinears in Bogoliubov coefficients are given by
equations analogous to \eq{a-b-ident}
\begin{align}
&|\alpha_{\rm eff}(k)|^2= \frac1{1- | \g_{\rm eff}(k)|^2},\; 
|\b_{\rm eff}(k)|^2= \frac{| \g_{\rm eff}(k)|^2}{1- | \g_{\rm eff}(k)|^2},\nonumber\\
&\alpha_{\rm eff}(k)\b_{\rm eff}^*(k)=  \frac{ \g_{\rm eff}(k)}{1- | \g_{\rm eff}(k)|^2},\; 
\alpha_{\rm eff}^*(k)\b_{\rm eff}(k)=  \frac{\g_{\rm eff}^*(k)}{1- | \g_{\rm eff}(k)|^2}
\label{a-b-sq}
\end{align}
where $\g_{\rm eff}$ is given by the equation \eq{gamma-eff}.

\subsection{Preparing an exact CC state}

We show here that although ground states of a massive theory under a
critical quench are given by a gCC state \eq{gCC}, with an infinite
number of $\k_n$ parameters (equivalently, an infinite number of
chemical potentials), by using the device we can explicitly
prepare an exact CC state.

We have seen that the squeezing function \eq{tailored} ensures that
$\k_n=0, n>4$.  Specializing to $\k_{4,0}=0, \k_{2,0}= \k_2$
leads to an exact CC state.
\begin{align}
| \psi, in \ran= | CC \ran \equiv  e^{-\k_2 H}| D\ran 
\label{tailored-kCC}
\end{align}
With the specific choice $\k_2 = 1/m_0$, the squeezing function and
the CC state reduce to
\begin{align}
& f_{m_0}(k)=1-\frac{2 |k|}{\sqrt{k^2+{m_0}^2} \tanh \left(|k|/m_0\right)+|k|}
\label{fCC}
\\
& | \psi, in \ran= e^{-H/m_0}| D\ran 
\label{tailored-CC}
\end{align}
We note that the squeezing function is a localized function which
vanishes at both $k\to0$ and $k\to\infty$ limits and hence the resultant
squeezed state is normalizable. Note that the functions $f(k)$ are
even functions, and hence are actually functions of $|k|$.

We can manufacture a CC state using an appropriate squeezing function
even when we consider a critical $\to$ critical quench. Applying the
squeezing method to the quench protocol discussed in Section
\ref{sec:critical}, we find that the following choice of the squeezing
function
\begin{align}
f(k)=\frac{a \left(-e^{2 |k| \left({\kappa_{2,0}}+{\kappa_{4,0}} k^2\right)}\right)+a-i |k|}{-a+
(a+i |k|) e^{2 |k| \left({\kappa_{2,0}}+{\kappa_{4,0}} k^2\right)}}\nonumber\
\end{align}
leads to a gCC state  $e^{-\k_{2,0}H-\k_{4,0} W_4}| D\ran$. Specializing to
\begin{align}
f(k)=\frac{a \left(-e^{2 {\kappa_{2,0}} |k|}\right)+a-i |k|}{-a+(a+i |k|) e^{2 
{\kappa_{2,0}} |k|}}\nonumber\
\end{align}
leads to a CC state $e^{-\k_{2,0}H}| D\ran$.

\section{\label{sec:fermion}Fermion theories with time-dependent mass}

We will now consider fermion field theories with a time-dependent mass:
\[
S= - \int d^2 x (i\bar\Psi \g^\mu \del_\mu \Psi - m(t) \bar\Psi \Psi)
\]
Once again, a general analysis of an auxiliary Schr{\"o}dinger problem can
be performed \cite{Duncan:1977fc}, to infer the emergence
of the general Calabrese-Cardy (gCC) state. However, we present below
the analysis for a specific mass quench protocol, involving a
$tanh$ function, which describes quantum quench from a non-critical to
a critical Hamiltonian.


\gap1

\noindent We start with the Dirac equation with the following
time-dependent mass:\cite{Duncan:1977fc, Das:2014hqa}
\begin{align}
m(t)=\frac{m_0}{2}\left(1-\tanh\left(\rho t\right)\right)\nonumber\
\end{align}
The Dirac equation is 
\begin{equation}
\left(i\gamma^\mu\partial_\mu - m(t)\right)\Psi=0 .
\end{equation}
The ansatz for a solution of this equation is
\begin{equation}
\Psi(k;x,t)=\left(\gamma^0\partial_t-\gamma^1\partial_x - i m(t)\right)e^{\pm ikx}\Phi(k,t)
\label{fantz}
\end{equation}
where $\Phi(k,t)$ is a two-component spinor that satisfies the following equation.
\begin{equation}
\left(\partial_t^2+k^2+m^2(t)-i\gamma^0\dot{m}(t)\right)\Phi(k,t)=0
\end{equation}
Defining $\Phi=(\phi_+,\phi_-)^T$, the equations decouple in the eigenbasis of $\gamma^0$ in the Dirac basis, 
\begin{equation}
\left(\partial_t^2+k^2+m^2(t)\mp i\dot{m}(t)\right)\phi_\pm(k,t)=0
\label{fdeom}
\end{equation}
where $\phi_+(t)$ is the solution corresponding to $\gamma^0$
eigenvalue $1$ and its part with asymptotic positive energy
eigenvalues appears with the spinor $u(0)$ in the mode expansion of $\Psi(x,t)$.
Similarly, $\phi_-(t)$ is the solution corresponding to $\gamma^0$
eigenvalue $-1$ and its part with asymptotic negative energy
eigenvalues appears with the spinor $v(0)$ in the mode expansion of $\Psi(x,t)$.
The conventions and the explicit solutions are described in Appendix
\ref{app:fermion}.  The explicit solutions lead to the following expressions of
Bogoliubov coefficients $\alpha_\pm(k)$ and $\beta_\pm(k)$.
\begin{align}
\alpha_\pm(k)&=\frac{\Gamma \left(-\frac{i|k|}{\rho}\right) \Gamma\left(1-\frac{i\omega_{in}}{\rho}\right)}{\Gamma \left(1-\frac{i \left(|k|\mp m_0+\omega_{in}\right)}{2\rho}\right) \Gamma \left(-\frac{i\left(|k| \pm m_0+\omega_{in}\right)}{2\rho}\right)}\\
\beta_\pm(k)&=\frac{\Gamma \left(\frac{i |k|}{\rho}\right) \Gamma \left(1-\frac{i\omega_{in}}{\rho}\right)}{\Gamma \left(-\frac{i \left(-|k|\pm m_0+\omega_{in}\right)}{2\rho}\right) \Gamma \left(1-\frac{i\left(-|k|\mp m_0+\omega_{in}\right)}{2\rho}\right)}
\label{dirac-bogo}
\end{align}
In terms of the `out' oscillators, the `in' ground state is
\begin{align}
 |\psi\rangle = \exp\left[\sum_{k=-\infty}^{\infty} \gamma(k) a^\dagger_{k,out} b^\dagger_{-k,out}\right] |0,out\rangle\nonumber\
\end{align}
where $\gamma(k)=\chi(k)\frac{\beta_+(k)^*}{\alpha_+(k)^*}$ (\ref{fermion-trans}).
Using a similar BCH formula to \eq{BCH} for fermionic creation and annihilation operators, we get
\begin{eqnarray}
\label{psigg} &&\quad|\Psi\rangle =e^{-\kappa_2 H+\kappa_4W_4-\kappa_6W_6-...}|D\rangle\\
\text{where} \quad&&\kappa_2=\frac{1}{2 m}+\frac{\pi ^2 m}{12 \rho ^2}+\frac{1}{m}\,\mathcal{O}(m/\rho)^3, \; \kappa_4=\frac{1}{12 m^3}-\frac{\pi ^2}{24 m \rho ^2}+\frac{1}{m^3}\,\mathcal{O}(m/\rho)^3,\nonumber\\
&&\kappa_6=\frac{3}{80 m^5}-\frac{\pi ^2}{96 m^3 \rho ^2}+\frac{1}{m^5}\,\mathcal{O}(m/\rho)^3\nonumber\
\end{eqnarray}
and $|D\rangle$ is the Dirichlet state of the fermionic theory. Using the chiral mode expansion (\ref{crep}) and (\ref{crepb}), 
\begin{eqnarray}
 |D\rangle=e^{\sum_k\text{sign}(k)a_k^\dagger b_{-k}^\dagger}|0\rangle.
\end{eqnarray}
In writing the ${\mathbb W}_\infty$
charges for the fermions, we have used the currents mentioned in the
Appendix \ref{app:fermion}.\footnote{We choose the overall
  normalization of the $W_{2n}(z)$-currents so that the $W_{2n}$
  charges are given by $W_{2n}= \sum_k
  |k|^{2n-1}\left[a^\dagger(k)a(k) + b^\dagger(k) b(k) \right]$.}

\section{Correlators\label{sec:corr}}

The purpose of this section is to explicitly compute Wightman
functions of the kind \eq{wightmann} to study their exact time
evolution.

We first review the calculation of correlation functions in a purely
thermal ensemble and GGE in section \ref{sec:GGE}. Next in section
\ref{sec:exact-gr}, we calculate the same correlation functions in the
ground state quench and show that it cannot be approximated by a CC
state (or thermal ensemble in the long time limit) because of its
infinite number of conversed charges. In section \ref{sec:exact-mu4},
we repeat the same exercise for a precisely prepared CC state and gCC
state with $W_4$ charge (as in section \ref{sec:squeeze}). In these cases,
we explicitly see thermalization of a CC state and gCC state to a thermal
ensemble and GGE respectively as expected from MSS. 

\subsection{\label{sec:GGE}Correlators in thermal ensemble and GGE}
\paragraph{Real time propagator in a thermal ensemble}

Consider the real-time, thermal Wightman propagator (see,
e.g. \cite{Festuccia:2006sa} for the various definitions of
propagators).
\begin{align}
G_+(x_1,t_1; x_2, t_2; \beta) &\equiv \frac1Z \Tr\left(e^{-\b H}
\phi(x_1, t_1)\phi(x_2, t_2)\right)
\nonumber\\
&= \frac1Z \sum_{\{N_n\}} 
\lan \{N_n\} | \phi(x_1) e^{-it H}\phi(x_2)  e^{-it H} e^{-\b H} | \{N_n\}\ran
\end{align}
By using the occupation number representation of the Hamiltonian, it
is easy to derive the following result ($r= x_1 - x_2, t= t_1-t_2$):
\begin{align}
G_+(x_1,t_1; x_2, t_2; \beta) 
= \frac12 \int \frac{dk}{2\pi}
\left[
G_+(k;\b) e^{i k r - i |k| t}
+ G_-(k;\b) e^{-i k r + i |k| t}
\right],\nonumber\\
G_\pm(k,\b)= 
\frac1{ |k|(\pm e^{\pm \b |k|}\mp 1)} 
\end{align}
For $t_1=t_2$, we have,
\begin{align}
G_+(x_1,t_1; x_2, t_2; \beta) 
&= \int \frac{dk}{2\pi}\frac{1}{4k}\left(\frac{e^{ikr}e^{\beta|k|/2}+e^{-ikr}e^{-\beta|k|/2}}{2}\right)\csch\left(\frac{\beta k}{2}\right)\nonumber\\
&=-\frac{1}{2\pi}\log\left[\sinh\left(\frac{\pi r}{\beta}\right)\right]\
\label{therm_prop}
\end{align}
The two-point function of $\del\phi$ is,\footnote{We define $\del= \frac12(\del_x - \del_t)$,
  $\bar\del= \frac12(\del_x + \del_t)$.} therefore,
\begin{align}
\frac1Z \Tr\left(e^{-\b H}
\del\phi(x_1, t_1)\del\phi(x_2, t_2)\right)
= \int \frac{dk}{2\pi}\frac{e^{ik(r-t)}k}{4}\left( \coth \left(\frac{\beta k}{2}\right)+ 1\right)
=- \frac{\pi}{4\b^2}\frac1{\sinh^2(\pi(r-t)/\b)}
\label{phi-phi-thermal}
\end{align}
which is the well-known result obtained from CFT techniques
 \cite{Mandal:2015jla}. 

From the propagator expression \ref{therm_prop}, it is also easy to compute the thermal two-point function of exponential vertex operators.
\begin{align}
\lan \exp[i q\phi(0,t)] \exp[-i q \phi(r,t)] \ran_\b
= \left[\csch\left(\frac{\pi r}{\beta}\right)\right]^{q^2/(2\pi)}\xrightarrow{r\gg\beta}\frac{1}{2}\,e^{-q^2r/(2\beta)}\
\label{thermal-2pt-exp}
\end{align} 
Note that this result agrees with the expected result
\cite{Mandal:2015jla} from CFT, $\exp[-2 \pi
  \Delta r/\b]$ with $\Delta= q^2/4\pi$ (see Appendix
\ref{app:boson}).

The energy density in a thermal ensemble is
\begin{align}
\frac{E}{L}=\frac{\pi}{6\beta^2}.
\label{eq:thermalH}
\end{align}

We will now define the Wightman function in a GGE in an analogous
fashion (for simplicity first we consider only one chemical potential
$\mu_4$):
\begin{align}
G_+(x_1,t_1; x_2, t_2; \beta, \mu_4) &\equiv \frac1Z \Tr\left(e^{-\b H - \mu_4 W_4}
\phi(x_1, t_1)\phi(x_2, t_2)\right)
\nonumber\\
&\equiv \frac1Z \sum_{\{N_n\}} 
\lan \{N_n\} | \phi(x_1) e^{-it H}\phi(x_2)  e^{-it H} e^{-\b H- \mu_4 W_4} 
| \{N_n\}\ran
\end{align}
By a simple evaluation, this turns out to be
\begin{align}
G_+(x_1,t_1; x_2, t_2; \beta,\mu_4) 
= \frac12 \int \frac{dk}{2\pi}
\left[
G_+(k;\b, \mu_4) e^{i k x - i |k| t}
+ G_-(k;\b, \mu_4) e^{-i k x + i |k| t}
\right],\nonumber\\
G_\pm(k;\b, \mu_4)= 
\frac1{ |k|(\pm e^{\pm (\b |k| + \mu_4 |k|^3)}\mp 1)} 
\label{phi-phi-GGE}
\end{align}
The holomorphic two-point function is now given by
\begin{align}
\frac1Z \Tr\left(e^{-\b H}
\del\phi(x_2, t_2)\del\phi(x_1, t_1)\right)
&= \int \frac{dk}{2\pi} \frac{e^{ik(r-t)}k}{4}\left( \coth \left(\frac{\beta k}{2}+\frac{\mu_4 k^3}{2}\right)+ 1\right)
\label{th-del-del}
\end{align}
Generalizing \ref{th-del-del}, the holomorphic two-point function in a
GGE with arbitrary number of $W$ charges is
\begin{align}
&\frac{1}{Z}\text{Tr}\left(e^{-\beta H-\sum_n\mu_nW_n}\partial\phi(r,t)\partial\phi(0,t)\right)
=\frac{1}{Z}\text{Tr}\left(e^{-\sum_k \mu(k) N(k)}\partial\phi(r,t)\partial\phi(0,t)\right)\nonumber\\
&=\int\frac{dk}{2\pi}\frac{e^{ikr}}{4}\left(|k|\coth\left(\frac
{\mu(k)}{2}\right)+k\right),
\quad \mu(k) \equiv \beta |k|+\sum_{n\, \text{even}}\mu_n|k|^{n-1}
\label{gen-GGE}
\end{align}

\subsection{\label{sec:exact-gr}Exact time-dependent correlators in 
quantum quench: starting from ground state}

In this section, we will consider the specific quench protocol
discussed in Section \ref{sec:sudden}.\footnote{Note that the
  quantities defined in Section \ref{sec:sudden} are obtained by a
  naive definition of the sudden limit \eq{naive-sudden}. As explained
  in Appendix \ref{app:sudden-limit}, although for $W_4$ and higher
  charges, this definition has be refined as in \eq{precise-sudden},
  for correlator calculations we can continue to use the naive
  definition.}  Using the general computation \eq{phi-phi} of the
propagator and the specific values \eq{ab-sudden} and
\eq{wave-sudden}, we find
\begin{align}
G_{q,0}(x_1,t_1;x_2,t_2)~~ \equiv~~ & \lan 0,in  | \phi(x_1, t_1) \phi(x_2, t_2) | 0,in \ran=\nonumber\\
\int \frac{dk}{2\pi} \Big{[}G_{q,0}(k)
\{\left(
2k^2+m_0^2
\right) & \cos(k(t_1-t_2))-m_0^2\cos(k(t_1+t_2))\}-\frac{1}{4k}\left(e^{ikx}-e^{-ikx}\right) 
\Big{]}e^{ik(x_1-x_2)}\
\label{phi-phi-sudden}
\end{align}
Note that
first term involves the combinations $t_1 + t_2$, which reflect
the fact that time-translation invariance is lost due to the
time-dependent perturbation. In the above expression
\begin{align}
G_{q,0}(k)= \frac1{4|k|^2 \sqrt{k^2+ m_0^2}}
\label{g-k}
\end{align}
is the significant part of the propagator.  Singularities
of this quantity in the $k$-plane are explained Figure
\ref{fig-poles}: these consist of a double pole at $k=0$ and two branch
points on the imaginary axis, at $k= \pm i m_0$.

After performing the Fourier transforms, the propagator is given by:
\begin{align}
&
G_{q,0}(x_1,t_1; x_2, t_2)
\nonumber\\
& =\frac{1}{16\pi}\left[G_{1,3}^{2,1}\left(\frac{m_0^2}{4} \left(x_2^+-x_1^+\right){}^2\left|
\begin{array}{c}
 \frac{3}{2} \\
 0,1,\frac{1}{2} \\
\end{array}
\right.\right)+G_{1,3}^{2,1}\left(\frac{m_0^2}{4} \left(x_2^- -x_1^-\right){}^2\left|
\begin{array}{c}
 \frac{3}{2} \\
 0,1,\frac{1}{2} \\
\end{array}
\right.\right)\right.\nonumber\\
&-G_{1,3}^{2,1}\left(\frac{m_0^2}{4} \left(x_2^--x_1^+\right){}^2\left|
\begin{array}{c}
 \frac{3}{2} \\
 0,1,\frac{1}{2} \\
\end{array}
\right.\right)-G_{1,3}^{2,1}\left(\frac{1}{4} m_0^2 \left(x_2^+-x_1^-\right){}^2\left|
\begin{array}{c}
 \frac{3}{2} \\
 0,1,\frac{1}{2} \\
\end{array}
\right.\right)\nonumber\\
&+4 K_0\left(m_0\left|x_2^--x_1^-\right|\right)+4 K_0\left(m_0\left|x_2^+-x_1^+\right|\right)+2 i \pi\,\left(\text{sgn}\left(x_2^--x_1^-\right)-\text{sgn}\left(x_2^+-x_1^+\right)\right)\Bigg{]}
\label{phi-phi-explicit}
\end{align}
where we have defined $x_i^{\pm} = x_i \pm t_i$, $i=1,2$. In the asymptotic limit, for $x_1-x_2=r$ and $t_1=t_2=t$, this becomes
\begin{align}
G_{q,0}(0,t;r,t)&=\frac{1}{8}\left(m_0 (2t-r)\right)+\frac{1}{8\sqrt{2\pi m_0}}\left(\frac{e^{-m_0 (2 t-r)}}{\sqrt{2 t-r}}+\frac{e^{-m_0(r+2 t)}}{\sqrt{r+2 t}}+\frac{2 e^{-m_0 r}}{\sqrt{r}}\right)+ ...&r<2t\nonumber\\
&=\frac{1}{8\sqrt{2\pi m_0}}\left(\frac{e^{-m_0(r-2 t)}}{\sqrt{r-2 t}}+\frac{e^{-m_0(r+2 t)}}{\sqrt{r+2 t}}+\frac{2 e^{-m_0r}}{\sqrt{r}}\right)
+...&r>2t
\label{phi-phi-asym}
\end{align}
The linear terms are dictated by the double pole at the origin of the
$k$-plane. These agree with the expressions obtained by
\cite{Sotiriadis:2010si} in the so-called deep quench limit (see
Section \ref{sec:non-wilson} for more details). The ellipsis represent
higher transients.

\paragraph{Correlators:}

\begin{itemize}

\item
Two-point functions of vertex operators $O_q = e^{i q \phi}$:
  The dominant behaviour in the IR limit is given by exponentiating the
  linear part in the above $\lan \phi \phi \ran$ propagator (after
  subtracting the coincident part). We get
\begin{align}
&\lan 0,in  | e^{i q\phi(0, t)} e^{-iq \phi(r, t)} | 0,in \ran
= e^{-\frac{q^2}{8}m_0 r}, \quad t>r/2
\label{vertex-op}
\end{align}
This agrees with the result in \cite{Sotiriadis:2010si}. The dominant
exponential is, again, given by the double pole at the origin of the
$k$-plane. As remarked in Figure \ref{fig-poles}, the thermal
correlator is also dominated by this double pole at the origin. It is
no surprise therefore that the above result \eq{vertex-op} exactly
agrees with the thermal result \eq{thermal-2pt-exp}, with the
identification $\b= 4\k_2 = 4/m_0$.

\item 
Two-point functions of the holomorphic operator: $O=\del\phi$
\flushleft\vbox{\gap{-3}
\begin{align}
&\langle 0,\text{in}|\partial\phi(x_1,t)\partial\phi(x_2,t)|0,\text{in}\rangle\nonumber\\
&=\int \frac{dk\,e^{ikr}}{2\pi}\left[\frac{2k^2+m_0^2}{8\sqrt{k^2+m_0^2}}+\frac{k}{4}\right]
\label{new}\\
&=-\frac{m^2_0}{8\pi}\, K_2(m_0 r)\xrightarrow{r\to\infty}-e^{-m_0r} \left(+\frac{m_0^{3/2} \sqrt{\frac{1}{r}}}{8 \sqrt{2 \pi }}
+\frac{15 \sqrt{m_0} \left(\frac{1}{r}\right)^{3/2}}{64 \sqrt{2 \pi }}+O\left[\frac{1}{r}\right]^{5/2}\right)
\label{dpdpgq}
\end{align}
}
where we have chosen $r= x_1- x_2>0$ (note that there is no
time-dependence for equal times in this case, as we expect for
holomorphic operators since these do not `see' the boundary that
represents the quench). 

{\it Note that the derivatives annihilate the double pole at the
  origin of the $k$-plane, hence the two-point function is dictated
  solely by the distant singularity. Consequently, the rate of
  fall-off is NOT universal} (see Section \ref{sec:non-wilson} for
further details).

\underbar{Comparison with GGE:}

It is easy to see that the expression \eq{new} matches exactly with
the GGE value \eq{gen-GGE} with the following identification of the
chemical potentials given by \eq{rigol-chem-pot} $4\k(k) = \mu(k)$,
where $\k(k)= \k_2 |k| + \k_4 |k|^3 + ...$ (with specific values as in
\eq{kappa-sudden}) and $\mu(k) \equiv \beta |k|+\mu_4|k|^3+....$. 
This follows from the fact that 
\begin{align}
\kern-20pt|k|\coth\f{\mu(k)}{2}
=|k|\coth\left(2\kappa(k)\right)=|k|\,\frac{1+\gamma(k)^2}{1-\gamma(k)^2}
=|k|\,\frac{\alpha(k)^2+\beta(k)^2}{\alpha(k)^2-\beta(k)^2}=\frac{2k^2+m_0^2}{2\sqrt{k^2+m_0^2}}\
\label{del-del-comparison}
\end{align}
where we have used \eq{gamma-k} and \eq{BCH}, with $\alpha(k)$ and
$\beta(k)$ real and given by \eq{ab-sudden} (this manipulation
is similar to that in \eq{rigol-chem-pot}).

\end{itemize}

\begin{itemize}

\item
Two-point functions $\lan \del \phi(x_1,t) \bar \del\phi(x_2,t)\ran$:
\begin{align}
&\langle 0,\text{in}|\partial\phi(x_1,t)\bar{\partial}\phi(x_2,t)|0,\text{in}\rangle
&=-\int \frac{dk}{2\pi}\,\frac{m_0^2\,e^{i k(r-2 t)}}{8(k^2+m_0^2)^{1/2}}
&=-\frac{m_0^2}{8\pi} \, K_0(m_0(|r-2 t|))
\label{del-bardel-sudden}
\end{align}

\item
One-point function $\lan \del \phi\ \bar \del \phi (x,t) \ran$
\begin{align}
&\langle 0,\text{in}|\partial\phi\ \bar{\partial}\phi(x,t)|0,\text{in}\rangle
=-\int \frac{dk}{2\pi}\,\frac{m_0^2\,e^{-i 2k t}}{8(k^2+m_0^2)^{1/2}}
=-\frac{m_0^2}{8\pi} \, K_0(2m_0 t)
\nonumber\\
&\xrightarrow{t\to\infty} -e^{-2 m_0 t} 
\left[
\frac{m_0^{3/2} \sqrt{\frac{1}{t}}}{16 \sqrt{\pi }}-
\frac{\sqrt{m_0} \left(\frac{1}{t}\right)^{3/2}}{256 \sqrt{\pi }}+O\left(\frac{1}{t}\right)^{5/2}
\right]
\label{del-bardel-sudden-one}
\end{align}

\item
We also present a calculation of the energy density.  In the  $t\to\infty$
limit,
\begin{eqnarray}
\frac{E}{L}=m_0^2/(8 \pi)\
\label{quench-gr-energy}
\end{eqnarray}
Note that it does not agree with \eq{eq:thermalH} with $\b=4/m_0$.  In
other words, the higher chemical potentials affect the asymptotic
energy density.

\end{itemize}

\subsection{Squeezed state quench leading to correlators in CC states and simple gCC
states}\label{sec:exact-mu4}

In this subsection we will compute the exact quench evolution starting
from the precise squeezed states \eq{tailored-gCC}.  As argued before,
the calculations in this way reduce to simple gCC states and CC
states.

The expression for the $\lan f |\phi(x_1,t_1) \phi(x_2,t_2)| f \ran$
propagator in a general squeezed state is given in \eq{phi-phi-sq},
\eq{a-b-sq} and \eq{gamma-eff}. In this
subsection we will apply this to compute  correlators in the particular
squeezed state $|f_4\ran$ of \eq{tailored-gCC}. Recall that 
this state was tailored to
produce a given real value of $\k_2 >0$ and $\k_4 \ne 0$ (with all other
$\k_n=0$). For brevity, we will use the notation $\k_2$ and $\k_4$ instead of
$\k_{2,0}$ and $\k_{4,0}$ which are the specific values used in
\eq{tailored-gCC}. We find
\begin{align}
~\kern-300pt \lan f_4|\phi(r,t)\phi(0,t) | f_4\ran&=\int\frac{dk}{2\pi}\frac{e^{ikr}}{2k} \left(\coth \left(2 k \left(\kappa_2 +\kappa_4  k^2\right)\right)-\cos (2 k t) \csch\left(2 k \left(\kappa_2 +\kappa_4  k^2\right)\right)\right)\nonumber\\
\label{sqdeldel}\lan f_4|\partial\phi(r,t)\partial\phi(0,t) | f_4\ran&=\int \frac{dk}{2\pi} \frac{e^{ikr}k}{4}\left( \coth \left(2k\kappa_2+2k^3 \kappa_4\right)+ 1\right)
\\
\lan f_4 |\partial\phi\bar{\partial}\phi(x,t) | f_4\ran&=
-\int \frac{dk}{2\pi} \frac{e^{-2ikt} k}{4}\ \csch\left(2\kappa_2 k + 2k^3 \kappa_4\right)
\label{sq-del-del}
\end{align}
The first two equations describe two-point functions with
$(x_1,t_1)=(r,t)$, $(x_2, t_2)= (0,t)$, whereas the third equation is
a one-point function at a point $(x,t)$ (which is independent of $x$
by translational invariance).

\paragraph{\underline{$\k_4=0$}:}

~\\

With $\kappa_4=0$, i.e., for the CC state \eq{tailored-kCC}, 
the integrals can be done exactly and energy density can also be calculated in closed form,
\begin{align}
\lan CC | \phi(0,t)\phi(r,t) | CC\ran
&  =\frac{\log \left(\frac{1}{2} \text{csch}^2\left(\frac{\pi  r}{4 \kappa_2 }\right) \left(\cosh \left(\frac{\pi  r}{2 \kappa_2 }\right)+\cosh \left(\frac{\pi  t}{\kappa_2 }\right)\right)\right)}{8 \pi }\\
\lan CC |\partial\phi(0,t)\partial\phi(r,t) | CC\ran
&  =-\frac{\pi \csch^2\left(\frac{\pi  r}{4 \kappa_2 }\right)}{64 \kappa_2 ^2}\\
\lan CC| \partial\phi\!\bar{\partial}\phi(x,t)| CC\ran
&  =-\frac{\pi} {64 \kappa_2 ^2}\  \text{sech}^2\left(\frac{2\pi t}{4 \kappa_2 }\right)\
\end{align}

Note that

\begin{enumerate}

\item
The above results verify those that have been obtained using the
Calabrese-Cardy ansatz, applying the techniques of boundary CFT
\cite{Calabrese:2006quench}. We have derived these results here in the
context of an actual quantum quench starting from an appropriate
initial state (which ensures a CC post-quench state \eq{tailored-kCC},
as argued before).

\item
The two-point function of the `holomorphic' derivative operator
$\partial\phi$, computed in the second line, is independent of the
time $t$. It shows instant thermalization to the thermal value
\eq{phi-phi-thermal} (to match the two expressions, we need to
identify $\beta=4\kappa_2$ and put $x=r, t=0$).\footnote{The
  generalization to non-zero time difference between the two
  $\del\phi$'s is straightforward and it continues to agree with
  \eq{phi-phi-thermal}.}

\item
The energy density is
\begin{align}
\frac{E}{L}&=\frac{\pi }{96 \kappa_2 ^2}\
\label{quench-energy}
\end{align}
which also agrees with the thermal energy density in
(\ref{eq:thermalH}) with $\beta=4\kappa_2$.

\end{enumerate}

\paragraph{\underline{${ \k_4\ne 0}$}:}

~\\

With non-zero $\kappa_4$, we go back to the formulae 
\eq{sqdeldel} and \eq{sq-del-del} for 
general $|f_4\ran$ state.\\ 

We first take up $\partial\phi\bar{\partial}\phi$.  The one-point
function \eq{sq-del-del} can be evaluated using contour
integration. Note that the cosech function has simple poles in the
$k$-plane at
\begin{align}
2 \kappa _2 k + 2 \kappa _4
k^3 =i \pi n.
\label{pole-location}
\end{align}
Thus, there are three simple poles for each integer $n$ (see Figure
\ref{fig-poles}), given by
\begin{align}
&k_1=\frac{-2\ 6^{2/3} \kappa_2+\sqrt[3]{6} \left(\sqrt{48 \kappa_2^3-81 \pi ^2 \kappa_4 n^2}+9 i \pi\sqrt{\kappa_4} n\right)^{2/3}}{6 \sqrt[3]{\sqrt{3} \sqrt{\kappa_4^3 \left(16 \kappa_2^3-27 \pi ^2\kappa_4 n^2\right)}+9 i \pi \kappa_4^2 n}}
\nonumber\\
&k_2= \frac{4 \sqrt[3]{-6} \kappa_2+i \left(\sqrt{3}+i\right) \left(\sqrt{48 \kappa_2^3-81 \pi ^2 \kappa_4 n^2}+9 i
   \pi  \sqrt{\kappa_4} n\right)^{2/3}}{2\ 6^{2/3} \sqrt[3]{\sqrt{3} \sqrt{\kappa_4^3 \left(16 \kappa_2^3-27
   \pi ^2 \kappa_4 n^2\right)}+9 i \pi  \kappa_4^2 n}}
\nonumber\\
&k_3= -\frac{\sqrt[3]{-1} \left(2 \sqrt[3]{-6} \kappa_2+\left(\sqrt{48 \kappa_2^3-81 \pi ^2 \kappa_4 n^2}+9 i \pi \sqrt{\kappa_4} n\right)^{2/3}\right)}{6^{2/3} \sqrt{\kappa_4} \sqrt[3]{\sqrt{48 \kappa_2^3-81 \pi ^2\kappa_4 n^2}+9 i \pi  \sqrt{\kappa_4} n}}\nonumber\\
\label{poles-exact}
\end{align}
For positive $t$, only the poles in the lower half plane will
contribute to the contour integral. In an expansion in small
$\tilde\k_4 \equiv \k_4/\k_2^3$, we get
\begin{align}
k_1&=\frac{i \pi}{2 \kappa _2}\left(n +\frac{\pi^2 n^3}{4}\tilde\k_4
+o(\tilde\k_4^2)\right)
\label{k1-series}\\
k_2&=\frac{i}{\kappa _2}
\left(\f1{\sqrt{\tilde\kappa_4}}-\frac{\pi  n}{4}-\frac{3 \pi^2  n^2}{32} \sqrt{\tilde\kappa_4}+o(\tilde\k_4)\right)
\nonumber\\
k_3&=-\frac{i}{\kappa _2}
\left(\f1{\sqrt{\tilde\kappa_4}}+ \frac{\pi  n}{4}-\frac{3 \pi^2  n^2}{32} \sqrt{\tilde\kappa_4}+o(\tilde\k_4)\right)
\nonumber
\end{align}
Note that for $n=0$, $k_1=0$ (this is, in fact, exactly true, as can
be seen from \eq{poles-exact}). Thus, there is a pole at $k=0$ which
cancels the $k$ in the numerator in \eq{sq-del-del}. The poles
corresponding to $k_3$ are in the upper half plane, hence they are
irrelevant. The poles $k_2$ are in the lower half plane, but, in the
perturbative regime $\tilde\k_4 \ll 1$, have very large negative
imaginary parts for any value of $n$ (note the leading
$1/\sqrt{\tilde\k_4}$). Hence after the contour integration they lead
to very fast transients ($\sim$ $\exp[-2\f{t}{\k_2 \sqrt{\tilde
      \k_4}}]$). At long times, the dominant contribution to the
contour integral comes from the pole nearest to the origin, 
{\it i.e.} from the pole $k_1$ for $n=-1$.

From the expansion of $\csch\left(2 \kappa _4 \left(k-k_1\right)
\left(k-k_2\right) \left(k-k_3\right)+i \pi n\right)$ around $k_1$, we
find the residue of the $\csch$ at this pole to be
\begin{equation}
-\frac{1}{2 \kappa _4 \left(k_1-k_2\right) \left(k_1-k_3\right)}
\end{equation}
This leads to the following expression at long times
\begin{align}
\lan f_4| \partial\phi\!\bar{\partial}\phi(x,t)| f_4\ran=
-\frac{\pi}{16\kappa_2^2} \left(1+ \f{\pi^2}{4} 
\tilde{\kappa}_4+ \f{3\pi^4}{16} \tilde{\kappa}_4^2+
o(\tilde{\kappa}_4^3)\right) 
\exp\left(-t \f\pi{\k_2} \left(1+\f{\pi^2}4 \tilde{\kappa}_4+
\f{3\pi^4}{16} \tilde{\kappa}_4^2+
o(\tilde{\kappa}_4^3)\right)\right) 
\label{del-bardel-mu}
\end{align}

Let us now take up the
two-point function of $\partial\phi$. Note that \eq{sqdeldel}, is independent of
time, indicating instant thermalization (as was the case for the
CC state) and the integral is exactly equal to that
found in the GGE \eq{th-del-del} where $\beta=4\kappa_2$ and
$\mu_4=4\kappa_4$. 

The actual computation of the integral follows along similar lines as
above. Here, the poles are the same as in \eq{sq-del-del}. The
relevant residue, from $\coth$ at $k_1$, is
\begin{equation}
\frac{1}{2 \kappa _4 \left(k_1-k_2\right) \left(k_1-k_3\right)}
\end{equation}
Thus the total residue is similar to the earlier case. The final
result is 
\begin{align}
\kern-20pt\lan f_4| \partial\phi\!\bar{\partial}\phi(x,t)| f_4\ran=
-\frac{\pi}{16\kappa_2^2} \left(1+ \f{\pi^2}{4} 
\tilde{\kappa}_4+ \f{3\pi^4}{16} \tilde{\kappa}_4^2+
o(\tilde{\kappa}_4^3)\right) 
\exp\left(-r \f\pi{2\k_2} \left(1+\f{\pi^2}4 \tilde{\kappa}_4+
\f{3\pi^4}{16} \tilde{\kappa}_4^2+
o(\tilde{\kappa}_4^3)\right)\right) 
\label{del-del-mu}
\end{align}
which shows an exponent which is half of the thermalization exponent
found above. This is in accordance with MSS. We now make a detailed
comparison.

\paragraph{Comparison with MSS:} 

~\\

We will now compare \eq{del-bardel-mu} and \eq{del-del-mu} with
corresponding results in MSS.  Using the $W_4$-charge of the operator
$\del\phi\!\bar\del\phi$, namely $q_4=3$, along with the values
$\beta=4\kappa_2$ and $\tilde{\kappa}_4=\tilde{\mu}_4$, we find that
the result \eq{del-bardel-mu} matches that of MSS exactly.  Note the
higher order terms such as $\tilde{\mu}_4^2 t$ in the exponent. That
there is an exponentiation of a power series in $\tilde\mu_4$ was
anticipated in MSS on the basis of perturbative Feynman diagrams and
we verify this here explicitly.

The result \eq{del-del-mu} for the two-point function of $\del\phi$
also exactly matches MSS result.

Note that we resorted to perturbative expansion in small $\tilde \k_4$
to evaluate the integrals and compare with MSS.  There is no
non-perturbative calculation of the above correlators in gCC states
using CFT technique or other tools. The best result available is the
MSS result which we have reproduced here for the case of free scalar
theory with mass quench.

\section{\label{sec:non-wilson}Thermalization}

In the previous two sections, we found that the exact correlators
show thermalization at late times. Here's a brief summary for
some specific observables.

\begin{table}[H]
\centering
\begin{tabular}{|c|c|c|c|}
\hline
& Ground state $|0,in\ran$ &   CC state $e^{-H_0/m_0}|D\ran$ & Thermal state
\\
\hline
$\lan \del\phi(0,t)\del\phi(r,t)\ran$&
$\sim e^{-m_0 r}/\sqrt{r}$& $\sim e^{-\pi m_0r/2}$&
$\sim e^{-\pi m_0r/2}$ 
\\
\hline
$\lan e^{iq\phi(0,t)} e^{-i q\phi(r,t)}\ran$
&$\sim e^{-q^2m_0 r/8}$& $\sim e^{-q^2m_0 r/8}$ 
& $\sim e^{-q^2m_0 r/8}$ 
\\
\hline
energy density $\lan H \ran$
&$m_0^2/(16 \pi) $& 
$\pi m_0^2/96$& $\pi m_0^2/96$
\\
\hline
$\lan \del\phi\bar\del\phi(0,t)\ran$
&$\sim e^{-2m_0 t}/\sqrt{t}$& 
$\sim e^{-\pi m_0t}$& 0
\\
\hline
\end{tabular}
\centering
\caption{\footnotesize The 2nd and 3rd columns give equal time
  correlators at late times for a mass quench \eq{tanh-sudden}; the
  4th column gives the same correlator (time-independent) in a thermal
  state with $\b=4/m_0$. In the 2nd-column the initial state is the
  ground state $|0, in\ran$; in the 3rd column, the initial state is a
  special squeezed state \eq{tailored-CC} which is of the
  Calabrese-Cardy form $e^{-H/m_0}|D\ran$. In the first two rows, we
  list two-point functions at separated points. In the 3rd row we list
  the asymptotic energy density. In the 4th row, we list the late time
  behaviour of a one-point function; the vanishing asymptotic value
  agrees with the thermal state -- but we compare here the exponential
  decay in time between the second and third columns. Note that the
  asymptotic values always agree between the CC state and the thermal
  state, but barring the case of the exponential vertex operator, the
  late time behaviour differs from the CC state, signifying non-trivial
  modification of the behaviour by the higher chemical potentials.}
\label{table-comparison} 
\end{table}
Besides this, we also find an exact agreement between $t\to \infty$
correlators in the gCC state \eq{tailored} and in the corresponding
GGE ({\it cf.}  equations \eq{sq-del-del} and \eq{th-del-del}) with
chemical potentials $\mu_n= 4\k_n$. The relaxation rate of one-point
functions is seen to exactly exponentiate (see \eq{del-bardel-mu}),
and its perturbation expansion in the higher $\k_n$ coefficients
agrees with the MSS value \eq{gamma}. We also found in the previous
two sections that generically GGE correlators (equivalently, late time
correlators in a gCC state) and thermal correlators (equivalently late
time correlators in a CC state), characterized by the same temperature
(equivalently same $\k_2$) are different, even at large distances
(e.g. $\k_4$ appears in the correlation length in \eq{del-del-mu}).

It is clear from the above discussion and Table \ref{table-comparison}
that while the fact of thermalization is true, the late time exponents
depend nontrivially on the higher chemical potentials (or higher
$\k_n$'s), even though these correspond to perturbation by irrelevant
operators in an RG sense.  In the next subsection we address this
issue of sensitivity to irrelevant operators in some detail.  In the
following subsection we will discuss a second (related) issue of 
memory retention by the equilibrium ensemble through the higher
chemical potentials.

\subsection{UV/IR mixing}

In this section we will discuss the issue of large distance/time
universality (or the lack thereof) in a critical quench. A useful
guide in this turns out to be the pole structure of the propagator
$\lan \phi(k) \phi(-k) \ran$, which is explained in Figure
\ref{fig-poles}.

\begin{figure}[h]
\begin{center}
\begin{tabular}{ccc}
\framebox{
\begin{minipage}{0.3\hsize}
\begin{center}
\vspace{2ex}
\includegraphics[height=7cm, width=5cm]{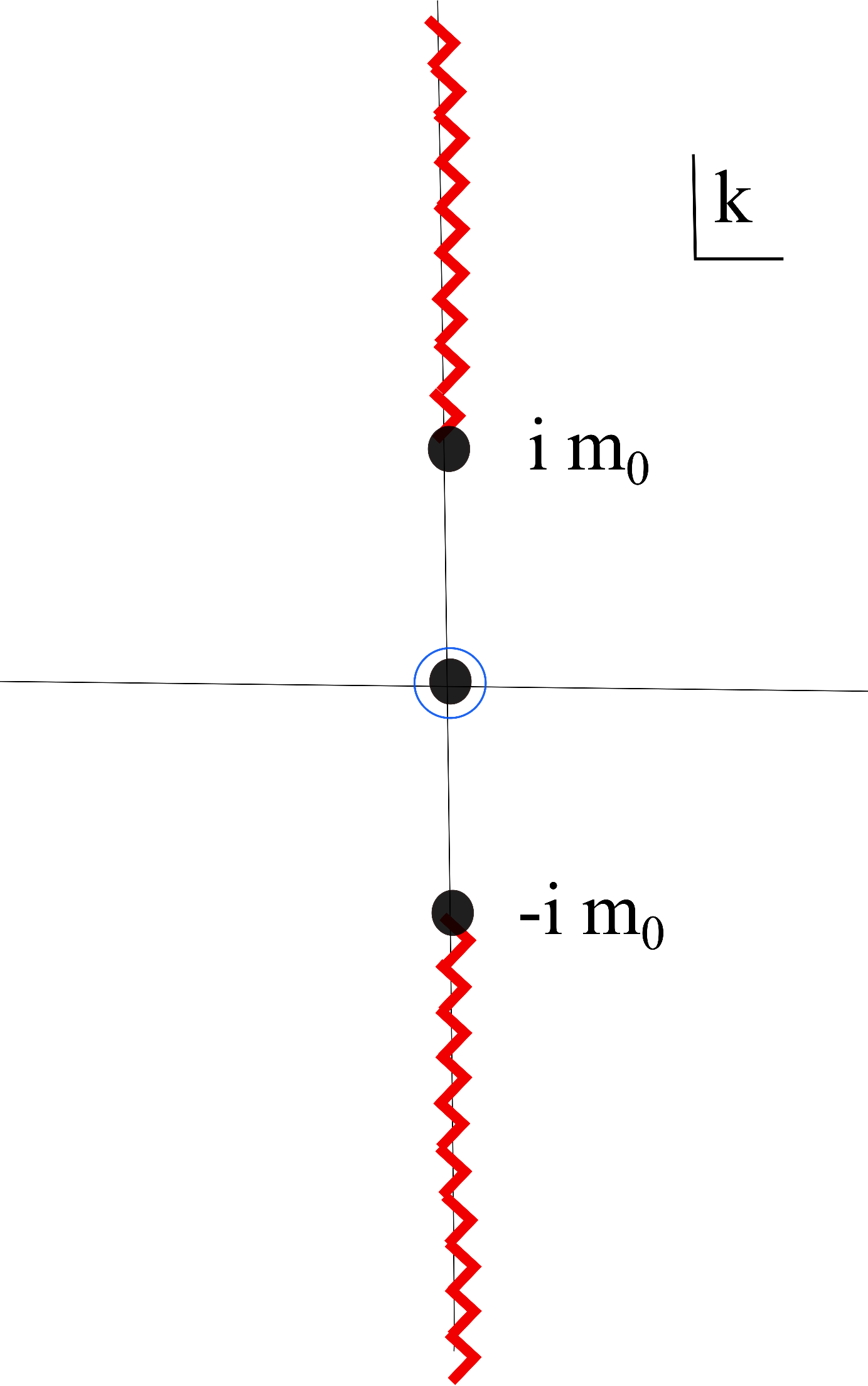}\\
{\bf (a)}
\end{center}
\end{minipage} 
}
\framebox{
\begin{minipage}{0.3\hsize}
\begin{center}
\includegraphics[height=7.35cm, width=5cm]{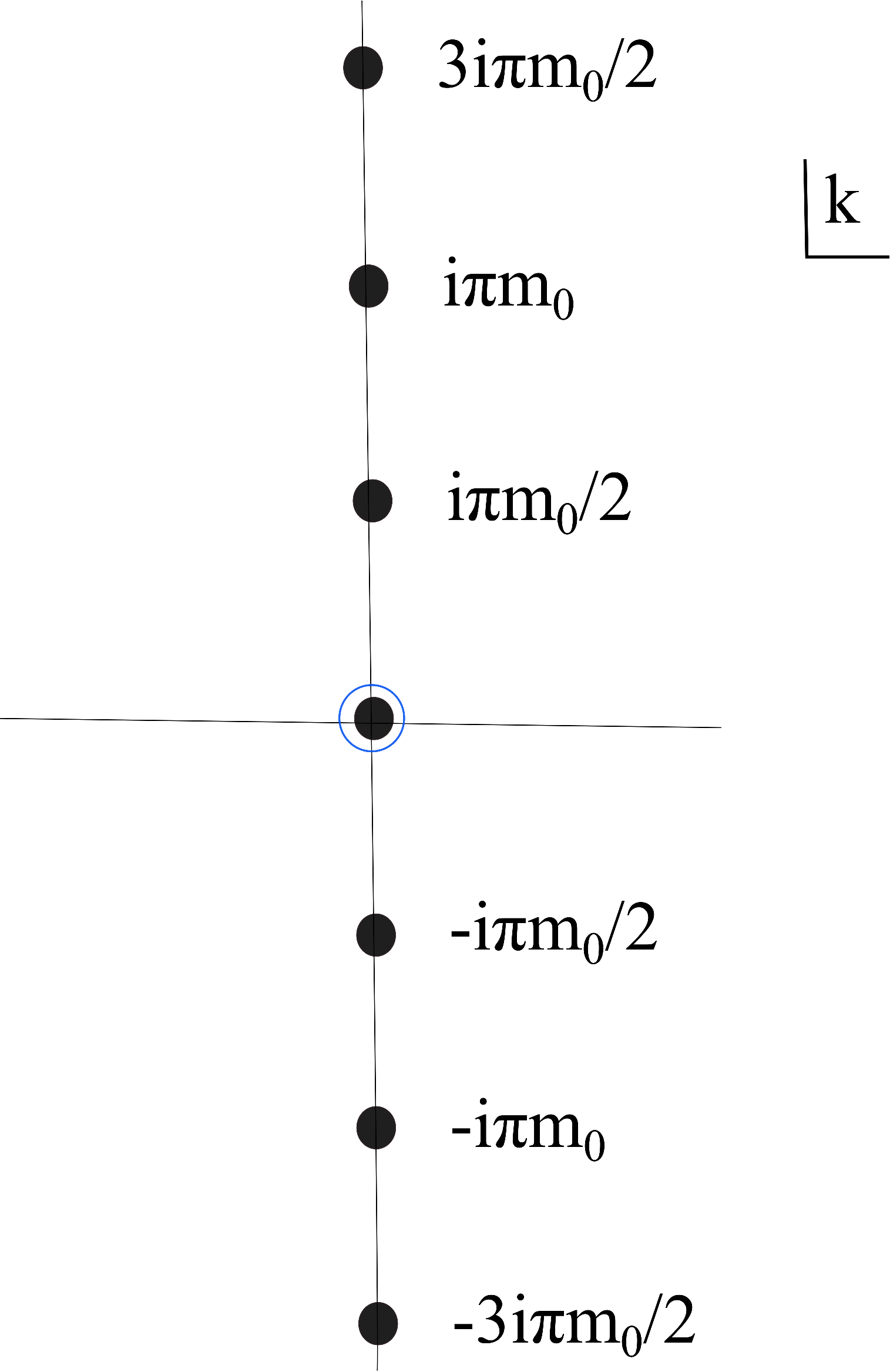}\\
{\bf (b)}
\end{center}
\end{minipage}
} 
\framebox{
\begin{minipage}{0.3\hsize}
\begin{center}
\vspace{2ex}
\includegraphics[height=7cm, width=3.5cm]{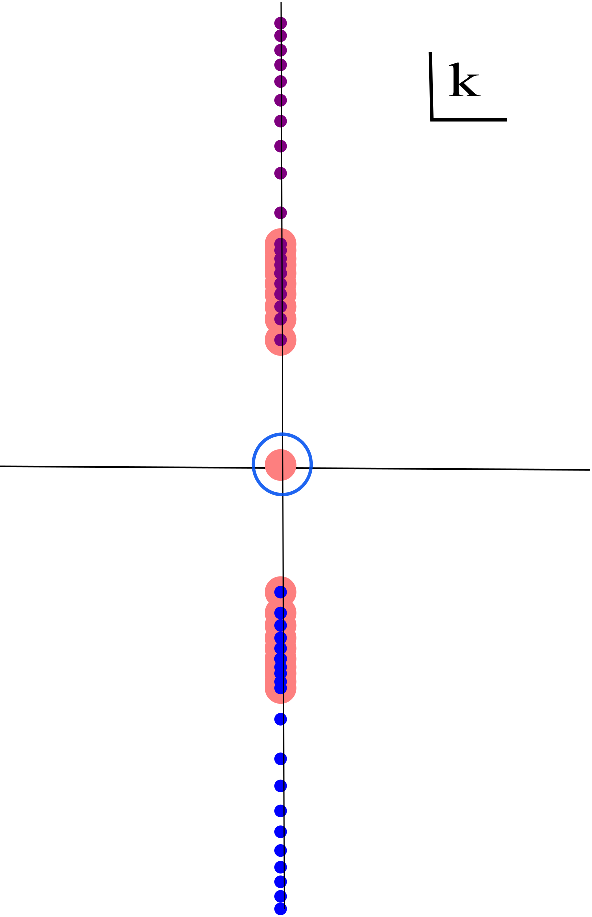}\\
{\bf (c)}
\end{center}
\end{minipage} 
}
\end{tabular} 
\caption{\footnotesize Singularities governing the two-point function
  in the complex $k$-plane: {\bf (a)} of the quantity $G_{q,0}(k)$ for
  the ground state quench propagator \eq{phi-phi-sudden}, {\bf (b)} of
  the quantity $G_\pm(k;\b)$ in the thermal propagator
  \eq{phi-phi-thermal}, {\bf (c)} of the quantity $G_\pm(k;\b, \mu_4)$
  in the GGE propagator \eq{phi-phi-GGE} with $\beta=2, \mu_4=0.2$; we
  have shown 30 leading poles. In each case the pole at the origin is
  a double pole, and yields the universal \underbar{linear large
    distance behaviour} of $\lan \phi\, \phi \ran$. Due to the
  equivalence between the quenched state and the gCC state
  \eq{ground-gCC}, the branch cut in (a) can be seen as a limiting
  case of an accumulation of single poles in a generalized version of
  (c) with an infinite number of chemical potentials determined by
  \eq{thermalization}, \eq{kappa-sudden}.  {\it In two-point functions such
    as $\lan \del\phi \del \phi\ran$, the double poles disappear and
    the large distance behaviour is sensitive to the sub-leading
    singularities, which are clearly different. This shows different
    types of large distance behaviour which are sensitive to the
    presence of higher dimensional operators.}}
\label{fig-poles}
\end{center}
\end{figure}

\paragraph{First look at universality:} 
~\\
Let us first discuss the naive argument
for universality in the present context. Note that in case of the
sudden quench we found \eq{gCC-sudden}
\[
|0, in \ran =  \exp[-\frac{H}{m_0}- \frac{5 W_4}{160 m_0^3} + ...]
| D \ran  
\]
which would appear to imply that, in the limit when the scale of the
quench is very high: $m_0 \to \infty$, the contribution of the
Hamiltonian is the most dominant and those of the higher dimensional
operators $W_{2n}$, $n>1$, are subdominant. This argument, of course,
is flawed, since $m_0$ is dimensionful, and the important issue is
the relative magnitude of $W_4/m_0^3$ {\it versus} $H/m_0$ in
a particular state.

There are, of course, more refined arguments for universality which
define an IR limit in terms of dimensionless distances and times
\begin{align}
m_0 r, m_0 t \gg 1
\label{deep-quench}
\end{align}
which is called the deep quench limit in \cite{Sotiriadis:2010si}.
Ref. \cite{Sotiriadis:2010si} argues that in this limit, the
propagator in \eq{phi-phi-sudden} is dominated by the leading
expansion of the integrand in $|k|/m_0$, which is given by a double
pole. From \eq{phi-phi-asym}, we find that the leading behaviour of
this propagator is indeed given by the linear term which is solely
determined by this double pole. We find that this double pole and the
consequent leading behaviour exactly coincides with that of the
thermal propagator \eq{phi-phi-thermal}. Indeed, all three forms of
the $\lan \phi\kern1pt\phi \ran$ propagators, the quenched one
\eq{phi-phi-sudden}, the thermal one \eq{phi-phi-thermal} and the GGE
one \eq{phi-phi-GGE}, coincide in the leading behaviour. Thus, the
higher order chemical potentials do not modify the leading
behaviour. Note, however, that the subleading behaviours are rather
different in the three cases: the exponents are different;
furthermore, in the quenched propagator there is a prefactor involving
a square root.

\paragraph{\label{wilson}Lack of Universality:}
~\\
The long-distance/time leading behaviour of the $\lan \phi \phi \ran$
propagator is, of course, a rather limited part of the story. Does the
above universality hold for correlators involving other operators, in
particular, various primary fields (recall that $\phi$ is not a
primary field)?

To address this issue, we consider one-point functions of primary fields of the
kind $O(z, \bar z)= \varphi(z)\varphi^*(\bar z)$ which have a decay
rate given by \eq{GGE}, \eq{gamma}. For the sudden quench discussed in
Section \ref{sec:sudden}, using \eq{kappa-sudden}, we find that the
fractional contribution of $W_{2n}$ to the relaxation rate \eq{gamma}
is determined by the dimensionless quantity
\[
\tilde\mu_n = \mu_n/\b^{n-1} \sim \frac1{m_0^{n-1}}/\left(\frac1{m_0}
\right)^{n-1}
\]
which is of order one!  What has happened is that, since the quench is
characterized by a single scale, the chemical potentials due to the
higher dimensional operators are determined by the same mass scale as
the temperature, thus the dimensionless contribution due to $W_{n>2}$
is necessarily of order one. In fact, we would expect this kind of
behaviour in any single-scale quench (we will make comments about
multi-scale quench shortly).

Indeed, we find in \eq{del-bardel-sudden} that the leading behaviour
of the one-point function of $\del\phi\bar\del\phi$ is not given by
the thermal value (nor with any finite number of chemical potentials).
This is best understood by looking at the Figure \ref{fig-poles}.  The
derivatives $\del, \bar\del$ kill off the double pole at the origin in
all three diagrams, leaving singularities away from the origin.  These
in Figure (a) differ from those in Figure (b) or in Figure (c).
Figure (c), if redone with infinite number of chemical potentials as
given by \eq{kappa-sudden}, reproduce the singularities of Figure (a).

{\it Thus, we find that} all {\it higher dimensional operators are equally
  important in determining the long time behaviour of this
  operator. This is also what we anticipated from the MSS expression
  for the relaxation rate, as explained above.}

The same story holds for two-point functions $\lan O(x_1, t_1) O(x_2,
t_2) \ran$. The exact quench computation, even in the deep quench
limit \eq{deep-quench} is not reproduced by the thermal result or any
finite number of chemical potentials. This can be explicitly seen for
$O= \del\phi$ in the previous two sections. We have also verified this
lack of universality for operators which are a composite of
`derivative' operators and exponential vertex operators, e.g.  $O=
\del \phi e^{i q \phi}$. Once again, the reason is the annihilation of
the double pole at the origin by these generic operators.

It is only the pure exponential vertex operators $O= e^{i q \phi}$
whose two-point functions \eq{vertex-op} respect universality in the
deep quench limit, that is they are reproduced by the thermal behaviour
(these operators do not annihilate the pole at the origin). 

\paragraph{Multi-scale quench:}
~\\ 
The discussion above was mainly centered on a sudden quench from
the ground state of a massive theory with mass $m_0$. By starting from
squeezed states, we can introduce multiple scales in the
problem. Thus, as discussed in Section \ref{sec:exact-mu4}, it is
possible to construct a quantum quench resulting in two parameters
$\k_2$ and $\k_4$ which are independent (in terms of the asymptotic
GGE, the chemical potential $\mu_4$ is not determined by $\b$). The
argument about contribution of all higher charges being of the same
order, therefore, does not immediately hold. The question then is:
suppose we hold $\k_4$ fixed and small; can we recover thermal
behaviour (or, behaviour as in a CC) more or more accurately for
sufficiently large times or distances? 

To answer this, let us analyze the pole location given by
\eq{pole-location} 
\[
2\k_2 k + 2\k_4 k^3 = i\pi n
\]
It would be tempting to argue that for sufficiently small $k$ the
cubic term can be ignored to any order of accuracy. However, once we
consider correlators which cancel the pole arising from $n=0$, the
correlators in Section \ref{sec:exact-mu4} do not receive any
contribution at all from small $k$ region. Hence the $\k_4$ modifies
the solution of this equation by $1 + o(\k_4/\k_2^3)$ (more precisely,
the pole is given by $k_1$ from \eq{k1-series} for $n=-1$). Hence the
exponent itself changes depending on $\k_4$, and there is no sense in
which the contribution of $\k_4$ can be made less significant at larger
times!

\paragraph{Conclusion:} Generically universality, as defined above,
is violated. Long time/distance behaviour is affected by perturbing
the initial state by higher dimensional operators. The discussions in
this subsection suggest that single scale quenches may generically
show such a lack of universality; the same appears to be true of
multiple scale quenches where the correlators are dominated by poles
at a finite distance from the origin. 

\subsection{\label{sec:memory}Memory retention}

In this section we will discuss the issue of non-standard
thermalization in the models studied where the equilibrium chemical
potentials allow a reconstruction of the quench protocol (completely
or partially depending on the situation).

\paragraph{Quench from a ground state:} 
Let us first consider the case of quenches from a ground state.  As is
clear from \eq{gamma-k} and \eq{BCH}, the $\k_n$-coefficients of the
gCC state \eq{ground-gCC} have a one-to-one relation to the reflection
amplitude $r(k)$ of the analogous potential scattering problem 
discussed in Appendix \ref{app:schro}. Now, it is well-known that the
potential of a one-dimensional Schr\"odinger problem
\cite{inverse-scattering:2008,Cohen-1985, Deift-1979}\footnote{We thank
  Basudeb Dasgupta for pointing out the reference
  \cite{inverse-scattering:2008} to us.} can be reconstructed from the
reflection amplitude $r(k)$. As explained above, the potential of the
scattering problem is specified by $m(t)$.  Hence it follows that
$m(t)$ can be reconstructed from $\k(k)$. This, in turn, means that
the $\mu_n$'s, which are just $4\k_n$ \eq{thermalization} ($\mu_2
\equiv \b$), carry complete knowledge of the quench protocol
$m(t)$. Thus, the equilibrium ensemble remembers the quench protocol!
As an example, the coefficients $\k_n$ in \eq{kappa-s} can be used to
determine the parameters $m_0$ and $\rho$ which specify the quench
protocol $m(t)$ completely.

\paragraph{Quench from excited states:} 
In case we consider a squeezed pre-quench state, there is additional
data in the pre-quench state and the GGE is characterized by the
function $\k_{\rm eff}(k)$ \eq{kappa-eff} which is given by a combination
of the knowledge of the squeezing function $f(k)$ and the quench
protocol $m(t)$ (see \eq{gamma-eff}).  For a known initial state,
the quench protocol $m(t)$ can be reconstructed from the GGE
parameters $\mu_n$ like above. Similarly, for a given quench protocol,
the initial state, characterized by $f(k)$ can be completely
determined by the $\k_n$-parameters (see, e.g.  \eq{tailored}). In
case the pre-quench initial state as well as the quench protocol are
unknown, the GGE allows reconstruction of a certain combination of the
data.

One might wonder whether, for a known quench protocol, any initial
state can be reconstructed from the final equilibrium ensemble. The
answer is no, as can be easily seen by considering a linear
combination of several squeezed states. In general, the final
equilibrium ensemble has only partial memory of the initial state.
Full reconstruction of the initial state happens only for special
states like the squeezed states considered in this paper. Of course,
besides the choice of the initial state, the integrability of the
model is another crucial ingredient for this result. We would comment
on the possible holographic interpretation of this result
in the next section.

\section{\label{sec:discussion}Discussion}

In this paper, we explicitly verify for actual critical quenches the
ansatz made in MSS for the generalized Calabrese-Cardy form (gCC)
\eq{gCC} of the initial state. We show that for an {\it arbitrary}
mass quench in a theory of free scalars as well as in a theory of free
fermions, a large choice of pre-quench initial states (ground state or
squeezed states) leads to a gCC state. We find that our results hold
even when the quantum quench begins and ends in a massless theory,
although in this case, the putative temperature sometimes turns out to
be imaginary and the issue of thermalization in these cases is subtle.

We find that while the ground state and generic squeezed states lead
to gCC states with all infinite number of $\k_n$ parameters present,
one can choose special squeezed states to prepare gCC states with
specific values of any given number of the $\k_n$-parameters; in
particular we can prepare a CC state of the form $e^{-\k_2 H}|D\ran$
from special squeezed states.

We compute the exact propagator in these quenches and hence the exact
time-dependence of correlators. We find that the correlators
thermalize at long times and the results verify those of MSS wherever
a comparison is possible. We have a simple understanding of the
identification \eq{GGE} of the $\k_n$'s with the chemical potentials
$\mu_n$ in terms of poles of the propagator. In specially prepared gCC
states with non-zero values of $\k_2$ and $\k_4$, we show that the
exponential decay given by the relaxation rate \eq{gamma} persists
non-perturbatively in $\k_4$.

We point out that the presence of the extra charges in the gCC state,
which are higher dimensional operators, non-trivially modify the long
distance and long time behaviour of correlators, in apparent
contradiction to Wilsonian universality. This is an example of a UV/IR
mixing; operators which are expected to be relevant in the UV by usual
RG arguments are found here to affect the IR behaviour of various
correlators. We present an understanding of this in terms of poles of
the propagator in the complex momentum plane. We find that while
exponential vertex operators do not suffer from these `non-universal'
corrections, all other operators (derivatives and composites of
derivatives and exponentials) do show this non-universal behaviour.

We also find another atypical behaviour, related to the above: the
equilibrium ensemble remembers about the quench protocol. In case
we start from the ground state of the pre-quench Hamiltonian, the
chemical potentials of the GGE encode a complete knowledge of
the quench protocol $m(t)$. With pre-quench squeezed state, the
chemical potentials encode a combination of information about
the initial state and the quench protocol.

\subsection{\label{sec:holography}Higher spin black holes}

In MSS \cite{Mandal:2015jla} we established a relation between (a)
relaxation of perturbations to a GGE in a CFT and, in the holographic
dual, (b) quasinormal decay to a higher spin (hs) black hole. In
particular, we found that the relaxation rate in (a) is equal to the
imaginary part of the quasinormal frequency \cite{Cabo-Bizet:2014wpa}
in (b). 

We also found in MSS that the rate in (a) is the same as the
asymptotic rate of thermalization to a GGE after a quantum quench.
This last result depended on an ansatz that the initial state is given
by a gCC state. In the present paper we have justified this ansatz; in
particular (see, e.g.  \eq{tailored-gCC}) we have shown explicitly
that a quantum quench from an appropriate squeezed state indeed leads
to such gCC states. In Section \ref{sec:exact-mu4} we have shown (see
\eq{del-bardel-mu}) that the exact formula for the relaxation rate
supports the perturbative formula \eq{gamma}.  Note that we now have
the relaxation rate non-perturbatively, including the two
non-perturbative branches \eq{poles-exact}. It would be interesting to
compare these two branches with the corresponding non-perturbative
branches of the hs black hole's quasinormal frequency
\cite{Cabo-Bizet:2014wpa}.

The above results prove the relation between quantum quench dynamics
in the field theory and quasinormal decay to higher spin black holes.
The specific hs black holes relevant to the present paper pertain to
the $\lambda\to 1$ and $\lambda\to 0$ limits of the
Gaberdiel-Gopakumar correspondence \cite{Gaberdiel:2010pz} in which
the dual conformal field theories describe free massless scalars and
free massless fermions respectively. The integrable structure
of the conformal field theories is reflected in the infinite
number of conserved charges of the hs black hole solutions.  

One may wonder if we can extend the above analysis to include
gravitational collapse to a hs black hole.  Note that a massive to
massless quench does not have a direct holographic dual since the
theory in the past is not conformal.  In this paper we have included a
brief discussion of quenches from a critical Hamiltonian to a critical
Hamiltonian, starting from ground states/excited states. This can
potentially describe a collapse geometry. The relevant CFT calculation
indicates that the quench history is determined in a one-to-one
fashion by the chemical potentials, or equivalently by the conserved
charges. This makes it plausible that in the process of gravitational
collapse to a hs black hole, the time-dependent history of the
`source' can be reconstructed from the final black hole configuration,
in a manner analogous to the dual CFT result on `memory retention'
(see \autoref{sec:memory}); the parameters specifying the
time-dependence are encoded in the infinite number of conserved
charges of the black hole.

\paragraph{Other open problems:}

Some of the obviously important extensions of the above work are to
the case of (i) massive to massive quenches, (ii) higher dimensions,
(iii) interacting theories. In particular, it would be interesting if
the phenomena of IR non-universality persists in higher dimensions.
The calculation of Bogoliubov coefficients and the exact propagator for
the tanh protocol appears to go through \cite{Das:2014hqa} in higher
dimensions in a straightforward manner. However, the analysis of the
poles requires more care. We hope to come back to this issue shortly.

\section*{Acknowledgement} We would like to thank Mustansir Barma, 
Basudeb Dasgupta, Kedar Damle, Deepak Dhar, Samir Mathur, Rob Myers,
Pranjal Nayak, Sreerup Raychaudhuri, Rajdeep Sensharma, Ritam Sinha,
Spenta Wadia and especially Sumit Das and Shiraz Minwalla for numerous
useful discussions. We would also like to thank the JHEP referee for
useful suggestions regarding reorganization of the sections and
improvement of clarity in the presentation. This work was partly
supported by Infosys Endowment for the study of the Quantum Structure
of Space Time.

\appendix

\section{\label{app:schro}The analogous scattering problem in 
quantum mechanics}

In the text, we encountered the Klein-Gordon equation \eq{eom1}
expressed in the form
\begin{align}
- \frac{d^2}{dt^2} \phi(k,t) + (k^2 -m^2(t)) \phi(k,t)=0
\label{eom1-app}
\end{align}
We will explain below an analogous Schr\"odinger problem and 
inferences for the solution of the  above problem. 

\subsection{Details of the quantum mechanics problem}

The equation above is analogous (see Table \ref{analogy-tab}) to the
Schr\"odinger equation for a particle in a potential (we will use the
convention $\hbar=1, 2m=1$)\footnote{We denote the spatial coordinate
of the analogous Schrodinger problem by $\tx$ to distinguish it
from the spatial coordinate $x$ of the original field theory problem.}
\begin{align}
-\ \frac{d^2}{d\tx^2} \psi(E,\tx) + (V(\tx)-E) \psi(E,\tx)=0
\label{map-app}
\end{align}
The dictionary is given by

\begin{table}[H]
\centering
\begin{tabular}{|c|c|}
\hline
Particle & QFT
\\\hline
$\tx$ & $t$
\\\hline
$E$ &  $k^2$
\\\hline
$V(\tx)$ & $-m^2(t)$
\\\hline
$\psi(E, \tx)$ &  $\phi(k,t)$
\\
\hline
\end{tabular}
\caption{Analogy}
\label{analogy-tab}
\end{table}
\noindent It is understood that $\phi(k,t)$
satisfies the reality condition $\phi^*(k,t)
= \phi(-k,t)$.

We will first discuss the quantum mechanical scattering problem. For a
mass-function $m^2(t)$ which drops from $m_0^2$ to zero, the potential
$V(\tx)$ asymptotes to $-U_0 \equiv - m_0^2$ as $\tx \to -\infty$ and
to zero as $\tx \to \infty$ (see Figure \ref{q-mech-fig}).  As is
well-known (see, e.g., \cite{landau1977quantum}, Section 25), the
wavefunction for such a problem takes the asymptotic forms
\begin{align}
\psi(\tx) & \xrightarrow{\tx\to-\infty}  A_1 e^{i k_1 \tx} + 
B_1 e^{-i k_1 \tx},
\;  k_1 =  \sqrt{(E+U_0)},
\nonumber\\
& \xrightarrow{\tx\to\infty}~ A_2 e^{i k \tx} + B_2 e^{-i k \tx}, \;
k  = \sqrt{E}.
\label{asym-psi}
\end{align}
The exact solution which matches the above asymptotic form with
$A_2=1, B_2=0$ is called the right-moving Jost function $f_+(k,\tx)$. There is another exact
solution which matches \eq{asym-psi} with $B_1=1, A_1=0$ which is
called the left-moving Jost function $\psi(k,\tx)= f_-(k,\tx)$. It can
be shown that (for generic momenta) $f_+(k,\tx)$ and $f_+^*(k,\tx)$
are independent, and so are $f_-(k,\tx), f_-^*(k,\tx)$. Clearly $f_-$
must be a linear combination of $f_+, f_+^*$ and {\it vice versa},
expressed in terms of Bogoliubov coefficients $\a_q, \b_q$
\footnote{The label `$q$', for `quantum mechanics', distinguishes
  these Bogoliubov coefficients from the analogous Bogoliubov coefficients $\a,
  \b$ that occur in the field theory discussion later on.}
\begin{align}
f_-(k, \tx)=  \a_q(k) f_+(k, \tx) + \b_q^*(k)  f_+^*(k, \tx)
\label{f-minus}
\end{align}
The existence of the Jost functions and properties relevant to the
present discussion can be proved under appropriate fall-off conditions
of the potential, see, e.g.  \cite{Cohen-1985, Deift-1979,
  inverse-scattering:2008}). In particular a sufficient condition
used in \cite{Cohen-1985} is
\begin{align}
\int_{-\infty}^\infty dx\, |V(x)- U_0
\theta(x)|(1+ |x|^2)< \infty
\label{fall-off}
\end{align}

\begin{figure}[H]
\centering
\includegraphics[scale=.35]{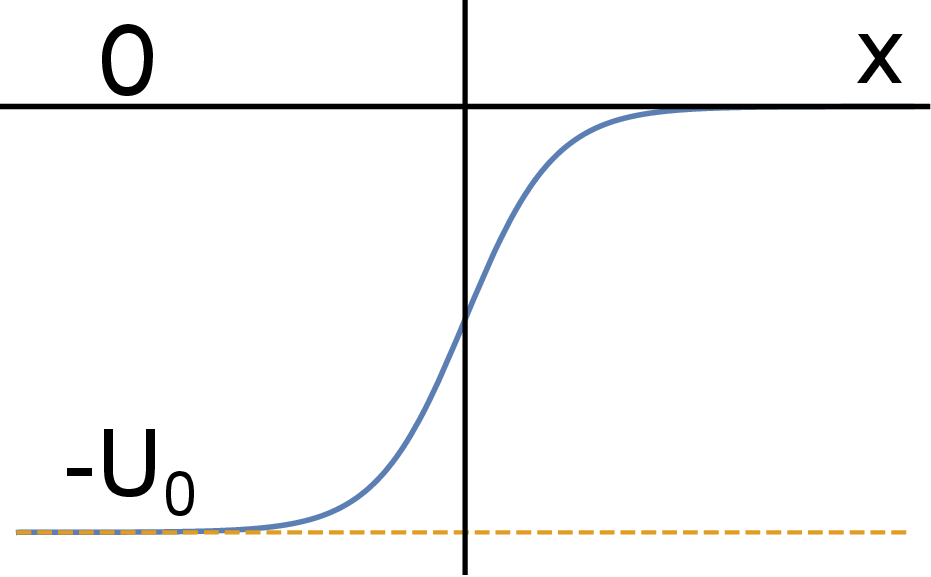}
\caption{\footnotesize Scattering of a particle in a potential $V(\tx)$.
$V(\tx) \to -U_0$ as $\tx \to -\infty$ and $\to 0$ as $\tx\to
\infty$}
\label{q-mech-fig}
\end{figure}

The general solution of the Schrodinger equation, which has
the asymptotic form \eq{asym-psi}, can be written in two alternative
forms
\begin{align}
\psi(k, \tx)=  A_1(k) f_-(k, \tx) +  B_1(k) f^*_-(k, \tx)
= A_2(k) f_+(k, \tx) +  B_2(k) f^*_+(k, \tx)
\label{qm-two-expansions}
\end{align}
By using \eq{f-minus} we find the following relations between
the  $(A_i, B_i)$  coefficients (called Bogoliubov
relations) \cite{landau1977quantum}:
\begin{align}
\left(\begin{array}{l}
A_2(k)\\ B_2(k)\end{array}\right)
=  \left(\begin{array}{ll}
\a_q & \b_q \\ \b_q^* & \a_q^*
\end{array}\right) 
\left(\begin{array}{l}
A_1\\ B_1\end{array}\right)
\label{q-mech-bogo}
\end{align}
The conservation of particle current along the positive
$\tx$-direction implies
\begin{align}
k_1 (|A_1|^2 - |B_1|^2) = k(|A_2|^2 -  |B_2|^2),
\label{continuity}
\end{align}
By applying \eq{q-mech-bogo}, we get
\begin{align}
|\a_q|^2 - |\b_q|^2= k_1/k
\label{bogo-norm}
\end{align}

Let us now consider a wave travelling from the left. This is
characterized by $B_2=0$ (i.e. as $\tx \to \infty$, $\psi$ becomes
purely right-moving).  The reflection and transmission amplitudes
$r\equiv B_1/A_1$ and $t \equiv A_2/A_1$ for such a wave can be easily
computed using \eq{q-mech-bogo}. We get
\[
r  =  -\b_q^*/\a_q^*, \;  
t=    \a_q + \b_q B_1/A_1= \a_q - |\b_q|^2/\a_q^*
\]
Note that for a wave of this kind the continuity equation
\eq{continuity} can be rewritten as
\begin{align}
k_1 (1- |r|^2)=  k |t|^2
\label{cont-2}
\end{align}
Similarly, for a wave travelling from the right, we must have $A_1=0$.
The corresponding reflection and transmission amplitudes $r'=
A_2/B_2$, $t'= B_1/B_2$ can be computed by using \eq{q-mech-bogo}:
\begin{align}
r' =  \b_q/\a_q^*,\;   t'= 1/\a_q^*
\label{rp-def}
\end{align}
It is clear that the two reflection amplitudes have the same
magnitude:
\[
|r'| = |r|
\]

\subsubsection{Power series expansion of the reflection amplitude}

It has been shown in \cite{Cohen-1985, Deift-1979,
  inverse-scattering:2008}, under conditions of sufficiently fast
fall-off of the potential function at $x= \pm \infty$
(\cite{Cohen-1985} uses \eq{fall-off}) that the right reflection
amplitude $r'$ has a Taylor expansion around $k=0$, with the leading
term $=-1$, thus:
\begin{align}
r'=  -1 + r_1 k + r_2 k^2 + ...,
\label{rp-expansion-app}
\end{align}
Below we will demonstrate some explicit examples of such an expansion.

\subsubsection{Examples of 1D scattering in QM\label{exPots}}

\begin{enumerate}
\item 
Consider a step potential $V(x)=-U_0\,\Theta(-x)$, with $U_0>0$.  
For such a potential, the asymptotic forms \eq{asym-psi} become
exact:
\begin{align}
\psi(\tx) & =  A_1 e^{i k_1 \tx} + 
B_1 e^{-i k_1 \tx}, \ \tx<0,
\;  k_1 =  \sqrt{(E+U_0)},
\nonumber\\
& = A_2 e^{i k \tx} + B_2 e^{-i k \tx},\ \tx>0, \;
k  = \sqrt{E}.
\label{exact-psi}
\end{align}
Demanding continuity of the wavefunction and of its first derivative
at $x=0$, we get
\[
\left(\begin{array}{l}
A_2\\ B_2\end{array}\right)=
\f12 \left(\begin{array}{ll}
1+ \f{k_1}{k} & 1- \f{k_1}{k}\\1- \f{k_1}{k} & 1+ \f{k_1}{k}
\end{array}\right) \left(\begin{array}{l} 
A_1\\ B_1\end{array}\right)
\]
Using \eq{q-mech-bogo} and \eq{rp-def} we can read off the following
Bogoliubov coefficients and reflection amplitudes
\[
\a_q=  \f12(1+ \f{k_1}{k}),\; \b_q= \f12(1- \f{k_1}{k}),\;
r'= \f{\b}{\a^*}= \f{1- \f{k_1}{k}}{1+ \f{k_1}{k}}, \;
r=- \f{\b^*}{\a^*}=- \f{1- \f{k_1}{k}}{1+ \f{k_1}{k}}, \;
k_1= \sqrt{k^2+ U_0}
\]
The reflection amplitude $r'(k)$ has a Taylor expansion around $k=0$:
\[
r'(k)=  
-1+\frac{2 k}{k_1}-\frac{2 k^2}{k_1^2}+O\left(k^3\right)
\]
which has the advertised form \eq{rp-expansion-app}.

\item Let us consider a piecewise continuous potential 
\[
V(\tx)= \left\{ \begin{array}{l} -U_0 , \; \tx \le 0\\
v,\; 0< \tx \le a\\
0, \;  \tx > a \end{array}\right.
\]
where $a>0$ and the constant $v$ can have any real value.  The
wavefunction has a piecewise form similar to \eq{exact-psi}, now in
three regions (with possibly complex momenta). By demanding continuity
of the wavefunction and of its derivative at $\tx=0$ and $\tx=a$, it
is straightforward to find {\footnotesize
\begin{align}
&\kern-40pt r'(k)
\nonumber\\
&\kern-50pt
=\frac{e^{-2 i a k} \left(\left(k \sqrt{-k^2+v}-\sqrt{-\left(k^2+U_0\right)
   \left(k^2+v\right)}\right) \cosh \left(a \sqrt{-k^2+v}\right)+i \left(-k
   \sqrt{k^2+U_0}+k^2-v\right) \sinh \left(a \sqrt{-k^2+v}\right)\right)}{\left(k
   \sqrt{-k^2-U_0+v}+\sqrt{-\left(k^2+U_0\right) \left(k^2-v\right)}\right) \cosh
   \left(a \sqrt{-k^2+v}\right)-i \left(k \left(\sqrt{k^2+U_0}+k\right)-v\right)
   \sinh \left(a \sqrt{-k^2+v}\right)}
\nonumber\\
&\kern-50pt
=-1+k\left[ 
\frac{2 
\left(-a v-i \sqrt{U_0}
\right) 
\sinh (a\sqrt v)+2 
\left(\sqrt v +i a \sqrt{U_0 v}
\right) 
\cosh(a \sqrt{v})
}
{
\sqrt{U_0 v} \cosh (a\sqrt v)+i v \sinh (a \sqrt v)
}\right]
+O\left(k^2\right).
\nonumber
\end{align}}
Note that the Taylor expansion of $r'(k)$ in the last line again
matches the form \eq{rp-expansion-app}.

\item For the potential $V(\tx) = -\f12 m_0^2 \left( 1 - \tanh(\rho
  \tx)\right)$, we find the following Bogoliubov
  coefficients
\begin{align}
\alpha_q(k) &= \frac{\Gamma \left(-\frac{i k}{\rho }\right) \Gamma \left(1-\frac{i k_1}{\rho }\right)}{\Gamma \left(-\frac{i (k+k_1)}{2 \rho }\right) 
\Gamma \left(1-\frac{i(k+k_1)}{2 \rho }\right)}, \quad
\beta_q(k) =\sqrt{\frac{k}{k_1}}\;  \frac{\Gamma 
\left(\frac{-i k}{\rho }\right) \Gamma \left(1-\frac{-i k_1}{\rho }
\right)}{\Gamma \left(\frac{-i (k - k_1)}{2 \rho }\right)
\Gamma \left(1+ \frac{-i (k- k_1)}{2 \rho }\right)}\nonumber\
\end{align}
We can compute these expressions by directly solving
the Schrodinger equation; here we have simply borrowed from
Section \ref{sec:tanh} and used the dictionary \eq{bogo-qm-qft}.
This leads to a reflection amplitude $r'(k)= \b_q/\a_q^*$
with the following Taylor series expansion:
\[
r'(k)=
-1+k \left(-\frac{2}{m_0}+\frac{-2 i \left(\gamma +
\psi ^{(0)}\left(-\frac{-i m_0}{2 \rho }\right)\right)}{\rho }\right)+O\left(k^2\right)
\]
which, again, agrees with the form \eq{rp-expansion-app}.

\end{enumerate}

\subsection{Implication for the field theory problem\label{app:implications}}

The Klein-Gordon equation for a massive scalar field $\phi(k,t)$ in
mixed Fourier space can be written as follows
\[
-\ \frac{d^2 \phi(k,t)}{dt^2} =(k^2+m^2) \phi(k,t)
\]
The classical solutions are well-known:
\[
\phi(k,t)= a(k) \frac{e^{-i \omega(k) t}}{ \sqrt{2\omega(k)}} + b(k)
\frac{e^{i \omega(k) t}}{ \sqrt{2\omega(k)}}, \; \omega(k)= + \sqrt{k^2 + m^2}
\]
If the scalar field is real, we have $\phi^*(k,t)= \phi(-k,t)$,
implying $b(k)= a^*(-k) $.

In the field theory problem, defined by \eq{eom1-app}, we have a mass
function $m^2(t)$ which asymptotes to $m_0^2$ and zero at $t\to \mp
\infty$. In analogy with the Jost function $f_-(k, \tx)$ introduced
above in the analogous Schr\"odinger problem, we must have an exact
solution $\phi(k,t)=u_{in}(k, t)$ which, in the infinite past,
asymptotes to
\[
u_{in}(k, t) \xrightarrow{t\to -\infty} \frac{e^{-i \omega_{in} t}}{
  \sqrt{2\omega_{in}}}, \; \om_{in}= \sqrt{k^2 + m_0^2}
\]
and an exact solution $u_{out}(k, t)$ (analogous to $f_+^*(k, \tx)$)
which asymptotes in the far future to
\[
u_{out}(k, t)\xrightarrow{t\to \infty} \frac{e^{-i \omega_{out} t}}{
  \sqrt{2\omega_{out}}}, \; \om_{out} = \sqrt{k^2}
\]
The normalization of the asymptotic wavefunctions are according to
standard conventions.  Just like \eq{qm-two-expansions}, we have two
alternative forms of the solution for $\phi(k,t)$:
\begin{align}
\phi(k,t)&= a_{in}(k) u_{in}(k,t) +  b_{in}(k) u^*_{in}(k,t) 
=a_{out}(k) u_{out}(k,t) +  b_{out}(k) u^*_{out}(k,t)
\nonumber\\
& \xrightarrow{t\to -\infty} 
\f{a_{in}}{ \sqrt{2\omega_{in}}} e^{-i \omega_{in} t} +  
\f{b_{in}}{ \sqrt{2\omega_{in}}} e^{i \omega_{in} t}
\nonumber\\
& \xrightarrow{t\to \infty} 
\f{a_{out}}{ \sqrt{2\omega_{out}}} e^{-i \omega_{out} t} +  
\f{b_{out}}{ \sqrt{2\omega_{out}}} e^{i \omega_{out} t}
\label{in-out}
\end{align}  
This can be compared with \eq{asym-psi} using Table \ref{analogy-tab}.
Further, using the reality properties $b(k)= a^*(-k)$ discussed
above, we get
\begin{align}
A_1(k) =  \f{a^*_{in}(-k)}{ \sqrt{2\omega_{in}}}, \;
B_1(k)= \f{a_{in}(k)}{ \sqrt{2\omega_{in}}}, \;
A_2(k) = \f{a^*_{out}(-k)}{ \sqrt{2\omega_{out}}}, \;
B_2(k)= \f{a_{out}(k)}{ \sqrt{2\omega_{out}}}
\label{comp-osci}
\end{align}
Just like in the quantum mechanics problem, $u_{in}$ is a linear
combination of $u_{out}$ and $u^*_{out}$ \cite{birrell1984quantum}
\begin{align}
u_{in}(k) =  \a(k) u_{out}(k) + \b(k) u^*_{out}(-k)
\label{u-in-u-out}
\end{align}
which leads to 
\begin{align}
a_{in}(k)= \a^*(k) a_{out}(k) + \b^*(k) a^*_{out}(-k)
\label{a-in-a-out}
\end{align}
Comparing with \eq{f-minus} and noting the extra normalization
factors $1/\sqrt{2 \om}$, we get
\begin{align}
\a(k) = \sqrt{\om_{out}/\om_{in}}~ \a_q(k),\;
\b(k) =\sqrt{\om_{out}/\om_{in}}~ \b^*_q(k)
\label{bogo-qm-qft}
\end{align}
Here we have used the fact that functions of $\om$ do not
distinguish between $k$ and $-k$.

This proves for us the important relation:
\begin{align}
\g \equiv \b^*/\a^* = \b_q/\a_q^*= r'
\label{gamma-rp}
\end{align}
Using the above result and \eq{rp-expansion-app}, we get a Taylor
series expansion
\begin{align}
\g = -1 + \g_1 k + \g_2 k^2 + \g_3 k^3 +  ...
\label{gamma-exp}
\end{align}
where the expansion coefficients are the same as in 
\eq{rp-expansion-app}, i.e. $\g_i = r_i$.

Another important relation, obtained from \eq{bogo-norm}
and \eq{bogo-qm-qft} is 
\begin{align}
|\a|^2 - |\b|^2= 1
\label{bogo-qft-norm}
\end{align}

\section{\label{app:BCH}Baker-Campbell-Hausdorff Calculation}

We will show that
\begin{align}
| \psi\ran \equiv
\exp\left(\frac12 \sum_k \g(k) a^\dagger (k)a^\dagger(-k)\right) |0 \ran
= \exp\left(-\sum_k \k(k) a^\dagger(k)a(k)\right) | Bd\ran
\label{gamma-kappa}
\end{align}
where\footnote{We thank Samir Mathur for drawing our attention to
  \cite{Mathur:1993tp} where a relation of the form \eq{kappa-def} was
  derived earlier in a somewhat different context for a single
  oscillator.}
\begin{align}
\k(k) =- \frac12 \log( \g(k)/\g_0)
\label{kappa-def}
\end{align}
and
\begin{align}
& |Bd\ran \equiv
\exp\left(\frac12 \sum_k \g_0 a^\dagger(k)a^\dagger(-k)\right) |0 \ran,
\label{bdry-state}
\end{align}
The choice $\g_0=-1$ corresponds to the Dirichlet state
\eq{dirichlet} (similarly, $\g_0=1$ corresponds to Neumann
boundary condition).  To derive \eq{gamma-kappa}, we note that the
right hand side can be written as
\[
\exp\left[\sum_k B(k)\right]\exp\left[\sum_k A(k)\right]| 0 \ran =
\exp\left[\sum_k B(k)\right]\exp\left[\sum_k A(k)\right] \exp\left[-\sum_k B(k)\right] | 0 \ran
\] 
where we have defined $B(k)=-\kappa(k)a^\dagger(k)a(k)$ and $A(k)=\gamma_0a^\dagger (k)a^\dagger (-k)$. The identity \eq{gamma-kappa} follows by noting that
 $[B(l),A(k)]=-\kappa(k)A(k)\left(\delta_{k,l}+\delta_{k,-l}\right)$, and by using the following form
of the  Baker-Campbell-Hausdorff (BCH) formula
\begin{equation}
e^Xe^Ye^{-X}=e^{\exp (s)Y}
\end{equation}
where $[X,Y]=sY$.

In the context of this paper, we will be interested in evaluating
$\k(k)$ from \eq{kappa-def} in a power series in $k$, using
\eq{gamma-exp}. Since the leading term in $\g(k)$ is $-1$, 
with the choice of the Dirichlet boundary state $\g_0=-1$, we
get the equation \eq{BCH} in the text. 

\section{\label{app:boson}Bosons}

The action for a free massless scalar is
\[
S = \frac12 \int dx dt \left[(\del_t \phi)^2 - (\del_x
  \phi)^2 \right]= - \frac12 \int dx dt\; \del_\mu \phi \del^\mu\phi
\]
The normal mode expansion is (we use ``box normalization'' $k=2\pi n/L$, $\int
\frac{dk}{2\pi}=\frac1L \sum\limits_n$)\footnote{We use the conventions of
  \cite{DiFrancesco:1997nk}.}
\begin{align}
\phi(x,t)&= \int \frac{dk}{2\pi} 
\left[\frac{a(k)}{\sqrt{2|k|}} \exp\left(ikx - i|k|t\right) 
+ \frac{a^\dagger(k)}{\sqrt{2|k|}} \exp\left(-ikx + i|k|t\right) \right]
\nonumber\\
&= \sum_{n\neq 0} \frac1{\sqrt{4\pi L |n|}} a_n  
\exp\left(\frac{2\pi}{L}(inx - i|n|t)\right) + {\rm h.c} 
\nonumber\\
&\equiv   \sum_{k\ne 0}
\left[\frac{a(k)}{\sqrt{2|k|}} \exp\left(ikx - i|k|t\right) 
+ \frac{a^\dagger(k)}{\sqrt{2|k|}} \exp\left(-ikx + i|k|t\right) \right]
\label{phi-normal-modes}
\end{align}
We will often use $a_n \equiv a(k)$, with a slight abuse of notation.
The commutation relations are $[a(k), a^\dagger(l)]= \delta_{kl}$.

\paragraph{Boundary states}

In terms of standard CFT oscillators $\a_n, \tilde\a_n$, the Dirichlet
boundary state is given by (see, e.g. \cite{Craps:2000zr} Eq. 4.1.13)
\[
| D \ran =  \exp[\sum_{n=1}^\infty \frac1n 
\alpha_{-n} \tilde\alpha_{-n}]|0\ran
\]
In terms of our oscillators $a_n \equiv a_k$
\begin{align}
& \alpha_{-n}= i \sqrt{n} a^\dagger_{-n},\;
\tilde \a_{-n}= i\sqrt{n} a^\dagger_n
\nonumber\\
&|D\ran = \exp[-\sum_{n>0} a^\dagger_n  a^\dagger_{-n}]|0 \ran
= \exp[-\frac12 \sum_{n\ne 0} a^\dagger_n  a^\dagger_{-n}]|0 \ran
= \exp[-\frac12 \sum_{k\ne 0} a^\dagger(k)  a^\dagger(-k)]|0 \ran
\label{dirichlet}
\end{align}
In the first step we used the relation between our oscillators here
and the standard CFT conventions (see \cite{DiFrancesco:1997nk},
Chap. 6).

\paragraph{Euclidean CFT}

We define $w= x+ i \tau$, $\bar w=x-i \tau$, $\tau=i t$. 
The Euclidean Propagator is
\[
\lan \phi(0,0) \phi(x,\tau) \ran=
\lan \phi(0,0) \phi(w,\bar w) \ran =  -\frac1{4\pi} (\ln w + \ln \bar w)
\]

\paragraph{Vertex operators}

Consider the exponential vertex operator $O(w, \bar w)= \exp[i
  q\phi(w,\bar w)]$.
\[
\lan \exp[i q\phi(0,0)] \exp[-iq \phi(w,\bar w)]\ran
=  w^{-q^2/4\pi} {\bar w}^{-q^2/4\pi}
\]
Hence $h=\bar h= q^2/8\pi, \; \Delta= q^2/4\pi$.
 
\paragraph{Boson W-currents}

We have used the following definitions of the ${\mathbb W}_\infty$
currents \cite{Bakas:1990ry} (normal ordering is implicit),
\begin{align}
T(z)&= \partial\phi(z)\partial\phi(z)\\
W_4(z)&= 2\partial^3\phi\partial\phi-3\partial^2\phi\partial^2\phi
\end{align}

\section{\label{app:fermion}Fermions}

We have used the following conventions in the text.
\begin{eqnarray}
&&\eta_{\mu\nu}= \begin{bmatrix}    1 & 0 \\ 0 & -1  \end{bmatrix}, \quad \partial_{\mu} = (\partial_t, \partial_x), \quad \gamma^\mu\partial_\mu=\gamma^0\partial_t-\gamma^1\partial_x,\nonumber\\
&&\gamma^0_d=\begin{bmatrix}    1 & 0 \\ 0 & -1  \end{bmatrix}, \quad\gamma^1_d=\begin{bmatrix}    0 & 1 \\ -1 & 0  \end{bmatrix}, \quad\text{in Dirac basis}.\nonumber\\
&&S=\frac{1}{\sqrt{2}}\begin{bmatrix}    1 & -1 \\ 1 & 1  \end{bmatrix}, \gamma^0_c=S\gamma^0_dS^{-1}=\begin{bmatrix}    0 & 1 \\ 1 & 0  \end{bmatrix}, \quad \gamma^1_c=S\gamma^1_dS^{-1}=\begin{bmatrix}    0 & 1 \\ -1 & 0  \end{bmatrix},\quad\text{in chiral basis.}\nonumber\\
&& u(0)= \begin{pmatrix} 1\\0\end{pmatrix},\qquad v(0)= \begin{pmatrix} 0\\1\end{pmatrix}\qquad \text{are the spinors in the rest frame.}\nonumber\
\end{eqnarray}
The spinors in a general frame are
\begin{align}
u(k,m)&=\frac{1}{\sqrt{(\omega+m)}}\begin{bmatrix} (\omega+m) \\ -k  \end{bmatrix}, & v(k,m)&=\frac{1}{\sqrt{(\omega+m)}} \begin{bmatrix} k \\ -(\omega+m)  \end{bmatrix}\nonumber\\
\bar{u}(k,m)&=\frac{1}{\sqrt{(\omega+m)}}\begin{bmatrix} (\omega+m) & k  \end{bmatrix}, & \bar{v}(k,m)&=\frac{1}{\sqrt{(\omega+m)}}\begin{bmatrix} k & (\omega+m)  \end{bmatrix}\nonumber\\
\label{norspinor}
\end{align}
where we have used the normalization $\bar{u}(k,m)u(k,m)=-\bar{v}(k,m)v(k,m)=2m$.
In the chiral basis, the mode expansion in the massless limit is
\begin{eqnarray}
 \Psi_c(x,t)&=&S\cdot\Psi(x,t)=\frac{1}{\sqrt{2}}\begin{bmatrix} 1 & -1 \\ 1 & 1 \end{bmatrix}\cdot \int \frac{dk}{2\pi}\frac{1}{\sqrt{2}}\begin{bmatrix} a_k e^{-ik\cdot x}+\text{sgn}(k)b^\dagger_k e^{ik\cdot x} \\ -\text{sgn}(k) a_k e^{-ik\cdot x} -b^\dagger_k e^{ik\cdot x} \end{bmatrix}\nonumber\\
 &=& \int_{-\infty}^{\infty} \frac{dk}{2\pi}\frac{1}{2}\begin{bmatrix} (1+\text{sgn}(k))(a_k e^{-ik\cdot x}+b^\dagger_k e^{ik\cdot x}) \\ (1-\text{sgn}(k))(a_k e^{-ik\cdot x}-b^\dagger_k e^{ik\cdot x}) \end{bmatrix}\
 \label{psic}
\end{eqnarray}
Writing as $\psi(x,t)$ and $\bar{\psi}(x,t)$,
\begin{eqnarray}
\label{crep} \psi(x,t)&=&\int_0^{\infty} \frac{dk}{2\pi}(a_k e^{-ik\cdot x}+b^\dagger_k e^{ik\cdot x})\\
 \bar{\psi}(x,t)&=&\int_{-\infty}^0 \frac{dk}{2\pi}(a_k e^{-ik\cdot x}-b^\dagger_k e^{ik\cdot x})\
 \label{crepb}
\end{eqnarray}

\paragraph{Solution of Dirac equation and Bogoliubov coefficients}

Using the coordinate transformation $y=e^{-\rho t}$ and the ansatz, we get the following equation:
\begin{equation}
\phi_\pm''(y)+\frac{\phi_\pm'(y)}{y}+\phi_\pm(y) \left(\frac{k^2}{\rho^2y^2}+\frac{m_0^2 y^2 \pm 2 i m_0 \rho}{\rho^2\left(y^2+1\right)^2}\right)=0
\end{equation}

The `in' solutions are solutions which become plane waves in far past and the `out' solutions are solutions which become plane waves in far future.
Due to the explicit $i$ in the equation of $\phi_\pm$, the positive energy solutions $\phi_{\pm,in/out,p}(k,t)$ and the negative energy solutions $\phi_{\pm,in/out,p}(k,t)^*$ are related as
\begin{align}
\phi_{+,in/out,m}(k,t)=\phi_{-,in/out,p}(k,t)^*,\qquad\phi_{-,in/out,m}(k,t)=\phi_{+,in/out,p}(k,t)^*\nonumber\
\end{align}
So, the solutions can be written as
\begin{align}
 \phi_{+,in/out}(k,t)&=\phi_{+,in/out,p}(k,t)+\phi_{-,in/out,p}(k,t)^*\nonumber\\
 \phi_{-,in/out}(k,t)&=\phi_{-,in/out,p}(k,t)+\phi_{+,in/out,p}(k,t)^*\nonumber\
\end{align}
The explicit solutions are
\begin{align}
\phi_{+,in}(k,t)&= \left(e^{-2\rho t}+1\right)^{-\frac{i m_0 }{2\rho}} e^{i t \left(\omega_{in}+m_0\right)} \,_2F_1\left(\frac{i\left(k-m_0-\omega_{in}\right)}{2\rho},\frac{i\left(-k-m_0-\omega_{in}\right)}{2\rho};1-\frac{i\omega_{in}}{\rho};e^{2\rho t}\right)\nonumber\\
\phi_{-,in}(k,t)&= \left(e^{-2\rho t}+1\right)^{\frac{i m_0}{2\rho}} e^{-i t \left(\omega_{in}-m_0\right)} \,_2F_1\left(\frac{i\left(k+m_0-\omega_{in}\right)}{2\rho},\frac{i\left(-k+m_0-\omega_{in}\right)}{2\rho};1-\frac{i\omega_{in}}{\rho};e^{2\rho t}\right)\nonumber\\
\phi_{+,out}(k,t)&=e^{-ikt}\left(e^{-2\rho t}+1\right)^{-\frac{im_0}{2\rho}} \,_2F_1\left(\frac{i\left(k-m_0+\omega_{in}\right)}{2\rho},\frac{i\left(k-m_0-\omega_{in}\right)}{2\rho};1+\frac{ik}{\rho};-e^{-2\rho t}\right)\nonumber\\
\phi_{-,out}(k,t)&=e^{-i kt} \left(e^{-2\rho t}+1\right)^{\frac{i m_0}{2\rho}} \,_2F_1\left(\frac{i\left( k+m_0-\omega_{in} \right)}{2\rho},\frac{i \left(k+m_0+\omega_{in}\right)}{2\rho};1+\frac{ik}{\rho};-e^{-2\rho t}\right)
\label{ftoutsol}
\end{align}
Defining the Dirac spinors as
\begin{align}
 U_{in/out}(k,x,t) &=K_{in/out}\left(\gamma^0\partial_t-i k \gamma^1 - i m(t)\right)e^{ikx}\phi_{+,in/out,p}(k,t)u(0)\nonumber\\
 V_{in/out}(k,x,t) &=-K_{in/out}\left(\gamma^0\partial_t+i k \gamma^1 - i m(t)\right)e^{-ikx}\phi_{+,in/out,p}(k,t)^*v(0)\nonumber\
\end{align}
where $K_{in/out}= i\left(\frac{1}{\omega_{in/out}+m_{in/out}}\right)^{1/2}$. For constant mass, $U(k,x,t)=u(k,m)e^{-ik\cdot x}$ and $V(k,x,t)=v(k,m)e^{ik\cdot x}$ where $u(k,m)$ and $v(k,m)$ have been defined in (\ref{norspinor}). The mode expansion of $\Psi(x,t)$ in terms of in/out modes are
\begin{align}
 \Psi(x,t)=\int_{-\infty}^{\infty} \frac{dk}{\sqrt{2\omega_{in/out}}} \left[ a_{k,in/out} U_{in/out}(k,x,t)+b^\dagger_{k,in/out} V_{in/out}(k,x,t)\right]\nonumber\
\end{align}
Using properties of hypergeometric functions  \cite{Abramowitz}, 
the Bogoliubov transformations between `in' and `out' solutions are
\begin{align}
\phi_{+,in,p}(k,t)&=\alpha_+(k)\phi_{+,out,p}(k,t)+\beta_+(k)\phi_{-,out,p}(k,t)^*\nonumber\\
\phi_{-,in,p}(k,t)&=\alpha_-(k)\phi_{-,out,p}(k,t)+\beta_+(k)\phi_{+,out,p}(k,t)^*\nonumber\
\end{align}
Hence, the Bogoliubov transformations between the `in' and `out' operators are
\begin{align}
 a_{k,in}&=
\left(
\frac{\omega_{in}}{\omega_{out}}\right)^{1/2}\frac{K_{out}}{K_{in}}\left(\alpha_+(k)^*a_{k,out}-\chi(k)\beta_+(k)^*b^\dagger_{-k,out}
\right)
\nonumber\\
 b_{k,in}&=
\left(
\frac{\omega_{in}}{\omega_{out}}
\right)^{1/2}
\frac{K_{out}}{K_{in}}
\left(
\alpha_+(k)^*b_{k,out}+\tilde{\chi}(k)\beta_-(k)^*a^\dagger_{-k,out}
\right)
\label{fermion-trans}
\end{align}
where $\chi(k)=\tilde{\chi}(k)=\text{sgn}(k)$.
It is straightforward now to find the expressions for the Bogoliubov coefficients
which are reproduced in the text \eq{dirac-bogo}.

\paragraph{Fermion W-currents}

We have used the following definitions of the super-${\mathbb W}_\infty$
currents \cite{Bergshoeff:1990yd} (normal ordering is implicit),
\begin{align}
T(z)&=-\frac{i}{2}\left(\psi^*\partial\psi(z)-\partial\psi^*\psi(z)\right)\nonumber\\
W_4(z)&=\frac{4}{5}q^2\left(\partial^3\psi^*\psi(z)-9\partial^2\psi^*\partial\psi(z)+9\partial\psi^*\partial^2\psi(z)-\psi^*\partial^3\psi(z)\right)\nonumber
\end{align}

\section{\label{app:sudden-limit}Subtleties of the sudden limit}

In Section \ref{sec:tanh} we analyzed the behaviour of the quench
under the ``tanh'' protocol for large $\rho$ in a power series in
$m_0/\rho$. In particular, in Section \ref{sec:sudden}, we defined the
sudden limit as the limit \eq{naive-sudden}. In this section we will
give a more precise definition of this limit. In certain quantities,
like the number operator \eq{number-tanh} in Section \ref{sec:tanh}
and the propagator in Section \ref{sec:exact-gr} etc. the distinction
is not essential, but in general the naive limit entails UV
divergences. E.g. all $W$-charges, including the energy density, under
a naive $m_0/\rho$ expansion introduced in Section \ref{sec:tanh}
appear to have progressively higher UV divergences as one goes down
the order. To treat these divergences properly, let us first analyze
these. Later on, we will find that terms in this expansion can be
resummed to yield finite expressions, provided we define the
sudden limit by the equation \eq{precise-sudden}.

~\\
The energy density is
\begin{multline}
E/L = \frac1{2\pi} \int_{-\Lambda}^\Lambda dk |k| N_k
= m_0^2\Bigg(\frac{1}{8 \pi }-\frac{m_0^2}{32 \pi  \Lambda ^2}+
O\left(\frac{m_0}{\Lambda }\right)^4
-\frac{m_0^2}{\rho^2}\left[\frac{1}{48} \pi 
   \log \left(\frac{\Lambda }{m_0}\right) \right.
\\
\left.
+\frac{1}{96} \pi  \log (4)
+ \frac{\pi m_0^2}{192 \Lambda ^2}+O\left(\frac{m_0}{\Lambda
   }\right)^4\right]+O\left(\frac{m_0}{\rho }\right)^4
\Bigg)\nonumber
\end{multline}
where we have used the asymptotic number density 
\eq{number-tanh}, in an $m_0/\rho$ expansion: 
\begin{align}
N_k 
& = \frac{\left(k-\sqrt{k^2+m_0^2}\right)^2}{4 k \sqrt{k^2+m_0^2}}-
\left(\frac{m_0}{\rho}\right)^2 \frac{\pi ^2 m^2_0}{48 \left(k
   \sqrt{k^2+m^2}\right)}+O\left(\frac{m_0}{\rho }\right)^4
\nonumber\end{align}
The $W_4$ density is
\begin{align}
W_4/L & = \int_{-\Lambda}^\Lambda \frac{dk}{2\pi} |k|^3 N_k= 
m_0^4 \Bigg[ \frac{4 \log (\Lambda/m_0 )-3+\log (16)}{64 \pi }+
  \frac{m_0^2}{32 \pi \Lambda ^2}+O\left(\frac{m_0}{\Lambda }\right)^4
\nonumber  
\\ &+ \left(\frac{m_0}{\rho}\right)^2 \left(-\frac{\pi \Lambda
    ^2}{96 m_0^2}+\frac{1}{192} \pi (2 \log (\Lambda/m_0 )-1+\log
  (4))+\frac{\pi m_0^2}{256 \Lambda ^2}+O\left(\frac{m_0}{\Lambda
  }\right)^4\right) + O\left(\frac{m_0}{\rho}\right)^4 \Bigg]
\nonumber
\end{align}
The $W_6$ density is
\begin{align}
W_6/L=  \int_{-\Lambda}^\Lambda \frac{dk}{2\pi} |k|^5 N_k=& 
m_0^6\Bigg[ \left(\frac{\Lambda ^2}{32 \pi m_0^2 }+\left(\frac{\log \left(\frac{m_0}{\Lambda }\right)}{16 \pi }+\frac{1}{24 \pi }-\frac{\log (4)}{32
   \pi }\right)-\frac{15 m_0^2}{512 \pi  \Lambda ^2}+O\left(\frac{m_0}{\Lambda }\right)^4\right)+
\nonumber\\
&\;\qquad+\frac{m_0^2}{\rho
   ^2}\Bigg(-\frac{\pi\Lambda
   ^4}{192 m_0^4}+\frac{\pi  \Lambda ^2}{192 m_0^2}+\frac{1}{128} \pi  \log \left(\frac{m_0}{\Lambda }\right)\nonumber\\
   &\qquad\qquad-\frac{1}{256} \pi  \log
   (4)+\frac{7 \pi }{1536}-\frac{5 \pi m_0^2}{1536 \Lambda ^2}+O\left(\frac{m_0}{\Lambda }\right)^4\Bigg)+ O\left(\frac{m_0}{\rho}\right)^4\Bigg]
\nonumber
\end{align}

\subsection{Resumming the divergences}

It turns out that the terms with growing UV-divergences with growing
powers of $m_0/\rho$ can be resummed to the following form.
\\
Introduce the scaling functions
\[
E/L  =  m_0^2 {{\cal E}}(x,y),\quad W_4/L=  m_0^4 F(x,y),
\quad W_6/L= m_0^6 G(x,y),
\quad x= m_0^2/\rho^2, \;  y= m_0^2/\Lambda^2
\]
The leading singularities in the above expressions for the charges are captured by
\begin{align}
{{\cal E}}(x,y)& = \frac1{8 \pi}+\frac{\left(\frac{5 \pi ^2 x}{8}+y\right) \log \left(\pi ^2 x+y\right)}{60 \pi} +  \cdots =  \frac1{8 \pi}+ \cdots
\nonumber\\
F(x,y)
&= -\frac{\left(\log \left(\frac{2 \pi ^4 x^2}{5}+y^2\right)+\log \left(\pi ^2 x+y\right)\right) \left(40 \left(5y +3\right)+\pi ^4 x^2+20 \pi ^2 x\right)}{11520 \pi }+ \cdots
\nonumber\\
G(x,y) &=
\frac1{1536 \pi }\left(\frac{\pi ^4 x^2}{120}+\frac{1}{32} \pi ^2 x (9 y+4)+y^2+y+1\right) \left[8 \left(\frac{25}{\sqrt{\frac{26 \pi ^4 x^2}{3}+25 y^2}}+\frac{1}{\pi ^2 x+y}\right) \right.
\nonumber\\
&\left.  + 19 \log \left(\frac{74 \pi ^4 x^2}{285}+y^2\right)+10
   \log \left(\pi ^2 x+y\right)\right]+ \cdots
\nonumber
\end{align}
~\\ The correct version of the ``sudden'' limit, therefore, is to take
the limit $\Lambda\to \infty$ first, for finite, large $\rho/m_0$ (see
Figure \ref{sudden-limit-fig}), i.e.
\begin{align}
y= \f{m_0^2}{\Lambda^2} \to 0, \; x= \f{m_0^2}{\rho^2}= {\rm small, ~fixed}
\label{precise-sudden}
\end{align}
In this limit, as we can see from the above
expressions:
\[
{{\cal E}}(x,0) = \frac1{8 \pi}+ \frac{\pi}{96} x \log(x) + \cdots
= \frac1{8 \pi}+ \cdots, \quad
F(x,0) \propto \log(x) + \cdots, \;  
G(x,0) \propto   \log(x)/x + \cdots, 
\]
which implies
\begin{align}
E/L  &= m_0^2\left( \frac1{8 \pi}- \frac{\pi}{48} \frac{m_0^2}{\rho^2} 
\log(\frac{\rho}{m_0})\right) + \cdots =
 m_0^2 \frac1{8 \pi} + \cdots  
\nonumber\\
W_4/L & \propto   m_0^4\  \log(\frac{\rho}{m_0}) + \cdots 
\nonumber\\
W_6/L &\propto   m_0^6 \  \frac{\rho^2}{m_0^2}  \log(\frac{\rho}{m_0})
 + \cdots
\end{align}

\begin{figure}[h]
\centering
\includegraphics[scale=.15]{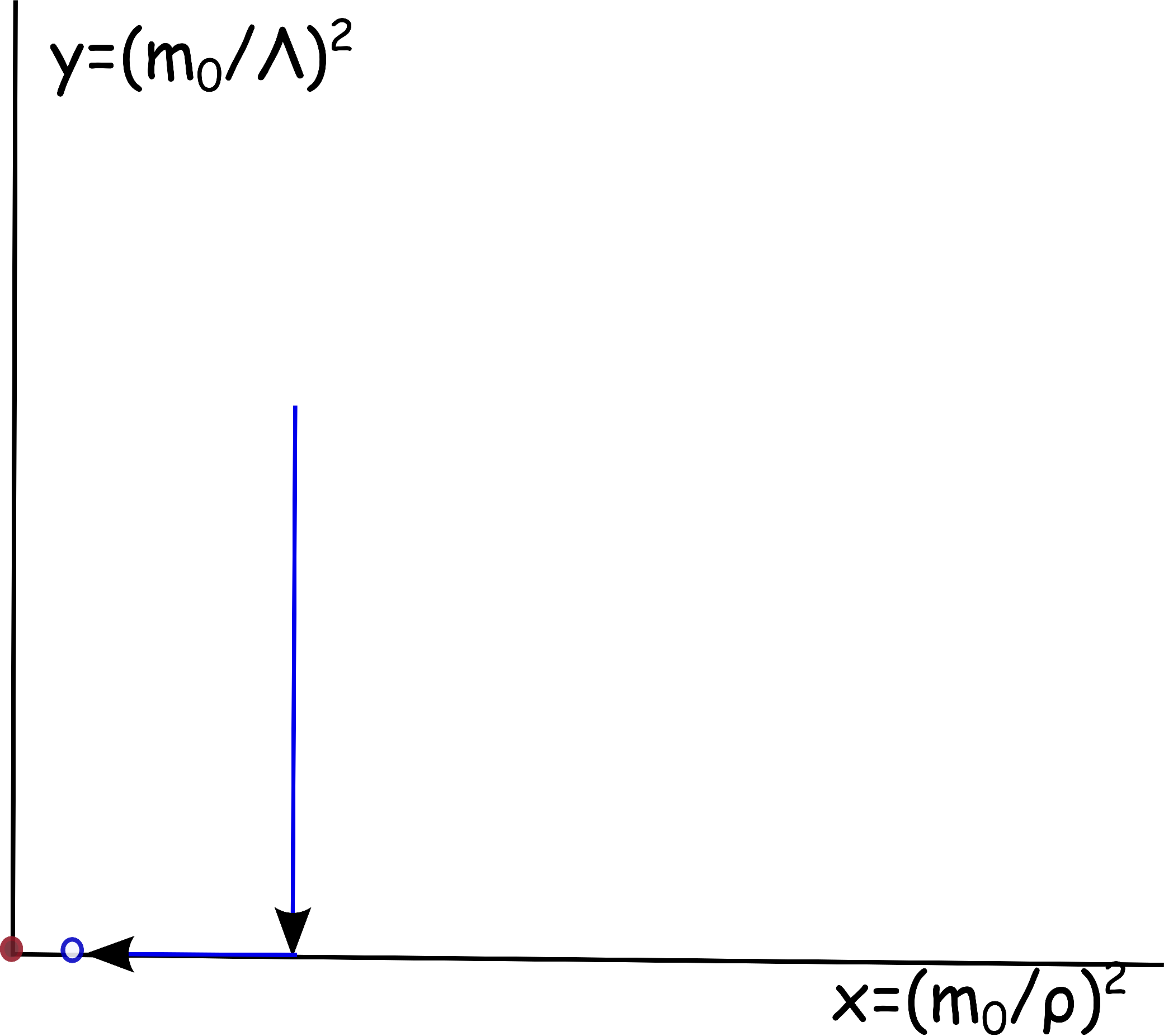}
\caption{The sudden limit.}
\label{sudden-limit-fig}
\end{figure}

\bibliographystyle{JHEP}
\bibliography{link-to-thermal-bib}

\end{document}